\documentclass[%
 reprint,
 superscriptaddress,
 amsmath,amssymb,
 aps,
 longbibliography,
]{revtex4-1}

\usepackage{graphicx}% Include figure files
\usepackage{dcolumn}% Align table columns on decimal point
\usepackage{bm}% bold math
\usepackage{float}
\usepackage{siunitx}
\usepackage{color}
\usepackage{verbatim}
\begin{document}

\preprint{APS/123-QED}

\title{Network architecture of energy landscapes in mesoscopic quantum systems} 
\author{Abigail N. Poteshman}
\affiliation{Department of Physics and Astronomy, University of Pennsylvania, Philadelphia, PA 19104, USA}
\author{Evelyn Tang}
\affiliation{Department of Bioengineering, University of Pennsylvania, Philadelphia, PA 19104, USA}
\author{Lia Papadopoulos}
\affiliation{Department of Physics and Astronomy, University of Pennsylvania, Philadelphia, PA 19104, USA}
\author{Danielle S. Bassett}
\email[To whom correspondence should be addressed: ] {dsb@seas.upenn.edu \& lbassett@seas.upenn.edu}
\thanks{These authors contributed equally.}
\affiliation{Department of Physics and Astronomy, University of Pennsylvania, Philadelphia, PA 19104, USA}
\affiliation{Department of Bioengineering, University of Pennsylvania, Philadelphia, PA 19104, USA}
\affiliation{Department of Electrical \& Systems Engineering, University of Pennsylvania, Philadelphia, PA 19104, USA}
\author{Lee C. Bassett}
\email[To whom correspondence should be addressed: ]{dsb@seas.upenn.edu \& lbassett@seas.upenn.edu}
\thanks{These authors contributed equally.}
\affiliation{Department of Electrical \& Systems Engineering, University of Pennsylvania, Philadelphia, PA 19104, USA}

\begin{abstract}
Mesoscopic quantum systems exhibit complex many-body quantum phenomena, where interactions between spins and charges give rise to collective modes and topological states. Even simple, non-interacting theories display a rich landscape of energy states\,---\,distinct many-particle configurations connected by spin- and energy-dependent transition rates. The collective energy landscape is difficult to characterize or predict, especially in regimes of frustration where many-body effects create a multiply degenerate landscape. Here we use network science to characterize the complex interconnection patterns of these energy-state transitions. Using an experimentally verified computational model of electronic transport through quantum antidots, we construct networks where nodes represent accessible energy states and edges represent allowed transitions. We then explore how physical changes in currents and voltages are reflected in the network topology. We find that the networks exhibit Rentian scaling, which is characteristic of efficient transportation systems in computer circuitry, neural circuitry, and human mobility, and can be used to measure the interconnection complexity of a network. Remarkably, networks corresponding to points of frustration in quantum transport (due, for example, to spin-blockade effects) exhibit an enhanced topological complexity relative to networks not experiencing frustration. Our results demonstrate that network characterizations of the abstract topological structure of energy landscapes can capture salient properties of quantum transport. More broadly, our approach motivates future efforts to use network science in understanding the dynamics and control of complex quantum systems. 
\end{abstract}

\maketitle
%\tableofcontents

\section{\label{sec:level1:Introduction} Introduction}
Mesoscopic quantum systems are open, many-body quantum systems that exist in the middle ground between microscopic quantum systems (\textit{e.g.}, individual atoms or electrons) and macroscopic phases of matter, such as superconductors and quantum Hall fluids \cite{kouwenhoven1997introduction}. Consisting of many (typically hundreds) of interacting particles, mesoscopic systems display complex many-body quantum effects and non-trivial energetics, including topological spin textures \cite{Gervais2005skyrmion,Gallais2008spinwave} and spin-charge separation \cite{Jompol2009spinchargeseparation}. Many well-known mesoscopic quantum systems\,---\,including quantum wires, quantum dots, and quantum antidots\,---\,exist in electronic systems with reduced dimensions, where they can be probed using quantum transport experiments. Electrons tunneling between the mesoscopic system and metallic reservoirs induce transitions between quantum-mechanical states, and these transitions can be detected through currents flowing in the device. Whether tunneling is allowed depends on the many-body configurations of the quantum system together with the spin and energy of electrons in the reservoirs.  

\begin{figure*}
  \centering
  \includegraphics[width=\textwidth]{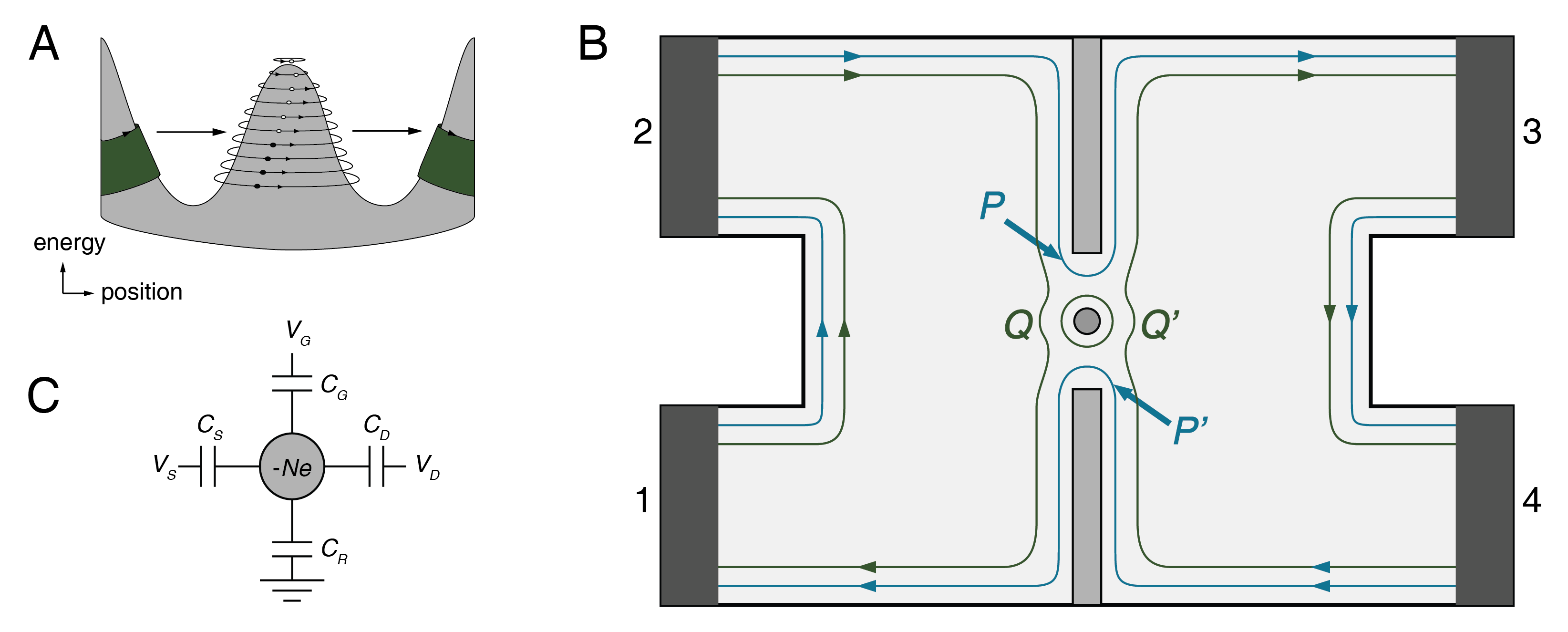}
  \caption{\textbf{An overview of the antidot device.} \textbf{A.} Schematic cross section of an antidot. Tunneling occurs between edge states carrying a sea of electrons (green) and the quantized antidot energy states. \textbf{B.} Schematic diagram of a typical antidot device displaying the edge-mode network for four-terminal conductance measurements of a single antidot. The modes of the upper edge that originate from contact 2 are split into $P$ modes that flow through the upper constriction and $Q$ modes that are reflected. Similarly, the lower modes that originate from contact 4 are split into $P'$ and $Q'$ modes that are transmitted and reflected, respectively. To measure the conductance, we apply a voltage to contact 1 to drive a current between contacts 1 and 3 $(I_{3})$, and we measure the potential difference between contacts 2 and 4 $(V_{24})$. The four-terminal conductance is given by $G = \frac{I_{3}}{V_{24}}$. \textbf{C.} Equivalent capacitor network for the antidot electrostatics. The quantized charge on the antidot is $-Ne$, where $N$ is the number of electrons (relative to a fixed reference configuration) and $e$ is the electron charge. The electrostatic potential of the antidot can be determined by the capacitive couplings ($C_S$, $C_D$, $C_G$) to the source, drain, and gate voltages ($V_S$, $V_D$, $V_G$). Any remaining coupling to other elements of the device is modeled as a capacitive coupling to the ground potential ($C_R$), such that the total capacitance is $C=C_S+C_D+C_G+C_R$. Figure adapted from \cite{bassett2009probing}.}
  \label{fig:ADdevice}
\end{figure*}

Here, we study transport through quantum antidots in the integer quantum Hall regime. A quantum antidot exists as a hill in the electrostatic potential landscape of a two-dimensional electron system; see Fig.~\ref{fig:ADdevice}A. Antidots exhibit discrete energy spectra due to magnetic confinement of electron orbital states, and can be treated as ``dots of holes,'' analogous to large quantum dots \cite{mace1995general,Sim2004kondo,Sim2008adreview,bassett2009probing}. In devices, quantum antidots can be coupled to extended, propagating edge modes of the integer quantum Hall fluid, where the tunneling to discrete antidot states is controlled by external voltages; see Fig.~\ref{fig:ADdevice}B. We consider a noninteracting model of antidot physics where a `state' refers to the specific way in which the $N$ holes in the antidot fill the single-particle energy levels \cite{mace1995general}, and transitions between states result from spin- and energy-dependent tunneling to the edge modes. Such a system can be described semiclassically, using a master equation to numerically simulate the effects of competing transport channels. However, the complexity of the interconnection patterns grows rapidly with $N$ (as for any quantum system, where the number of quantum eigenstates increases exponentially with the particle number), and it quickly becomes intractable to calculate the transport characteristics analytically for all but the simplest of systems.

\begin{figure*}
  \centering
  \includegraphics[width=\textwidth]{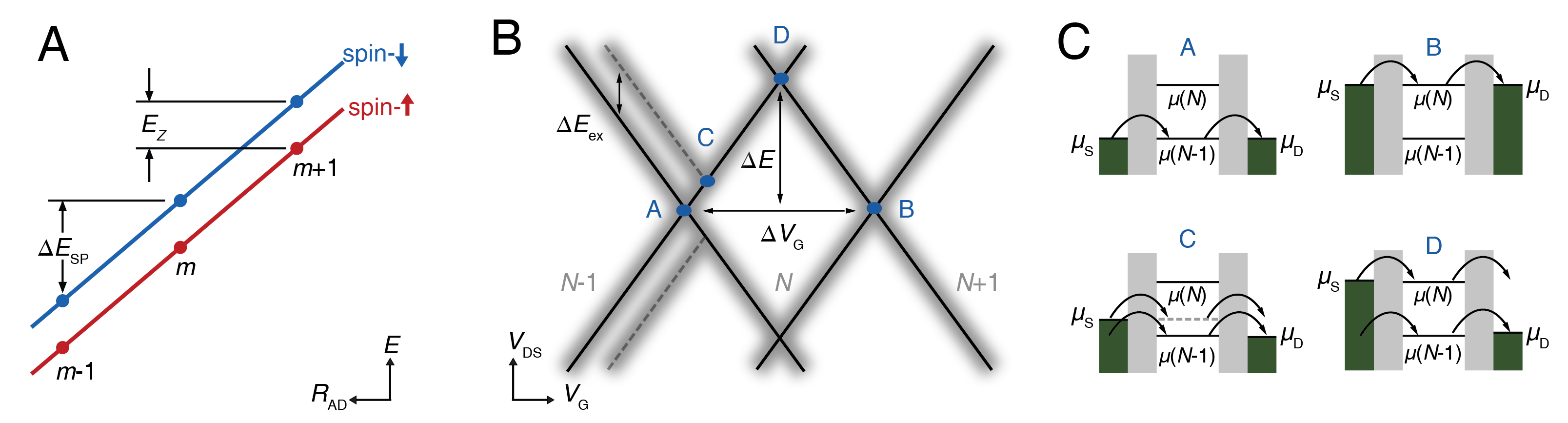}
  \caption{\textbf{An overview of antidot physics.} \textbf{A.} Single-particle energy spectrum of a $\nu_\mathrm{AD} = 2$ antidot in the space of energy $E$ versus radius of the antidot $R_{AD}$ where the spacing between the orbital energies $(\Delta E_{SP})$ is greater than the Zeeman energy $(E_{Z})$. In the single particle model, hole excitations can have the possible values $E_{ex} = \pm sE_{Z} + j\Delta E_{SP}$ where $s = 0$ or $s = 1$ represents spin-conserving or spin-flip transitions and $j$ is any integer. \textbf{B.} Schematic diagram of non-linear conductance through an antidot as a function of $V_G$ and $V_{DS}$. At the edges of the diamond, the electrochemical potential of the dot is aligned with either the source or the drain potential ($\mu_S$ or $\mu_D$) and the edges correspond to either the termination or the onset of quantum transport via tunneling. Inside the diamond, no transport occurs in a phenomenon known as a Coulomb blockade, and the number of electrons in the antidot is fixed. \textbf{C.} Schematic diagrams of the relative electrochemical potential of the antidot for different transitions and the transport window defined by $\mu_S$ and $\mu_D$, for several bias configurations as noted in panel \textbf{B}. Figure adapted from \cite{bassett2009probing}.}
  \label{fig:ADPhys}
\end{figure*}

\indent Network science provides a mathematical framework to characterize and describe complex interconnection patterns \cite{albert2002statistical,newman2010networks}. In this work, we use theories and computational tools from network science to characterize energy state transitions of quantum antidots containing many electrons. An increasing body of evidence suggests that complex systems with similar functions share organizational principles and converge to similar architectures \cite{barabasi2009scale}. One such organizational feature shared by many complex physical transport and information processing systems\,---\,including very large scale integrated computer circuits \cite{christie2000interpretation}, the vasculature system on the surface of rodent brains \cite{papadopoulos2018comparing}, large-scale human brain structural networks \cite{bassett2010efficient}, and the London Underground \cite{sperry2016rentian}\,---\,is a fractal network design known as Rentian scaling or Rent's rule. Rent's rule manifests as a scaling relation between the nodes and edges of a network, and reflects the presence of a hierarchical, self-similar architecture \cite{bassett2010efficient}. Furthermore, the topological Rent exponent can be used to quantify the degree of complexity of a network's interconnect topology, in that networks with higher topological Rent exponents have higher fractal dimensions.

\indent 
Although the energy landscapes of quantum antidots do not exist in physical space in the same way that computer circuits, biological tissues, and subway systems do, we hypothesize that the network architecture describing transport through quantum antidots may display similar patterns. Here we test this hypothesis by studying network representations of quantum transport, where nodes and edges represent energy states and available transitions, respectively. We find that the networks do exhibit Rentian scaling, and moreover that the degree of complexity correlates with key qualitative features of the transport models. The parallels to well-known network systems are illuminating; even though the energy-state networks have not been shaped by evolutionary or economic forces, they do represent a transport system that is constrained by physical laws, including energy and spin conservation. More broadly, our work demonstrates that a network-based approach to examining state transitions in physical systems can provide insights into the underlying physics that are currently inaccessible via other methods.

\section{\label{sec:level1Methods} Network models of quantum transport}

\subsection{\label{sec:level2:PhysChar} Physical characteristics of an antidot}

For an overview of quantum transport through mesoscopic systems generally and through antidots specifically, we direct the reader to reviews by other authors \cite{kouwenhoven1997introduction,Sim2008adreview}. Here, we focus on a single antidot in the integer quantum Hall regime at relatively low magnetic fields ($\begin{bf}B\end{bf}<$ \SI{1}{\tesla}) and antidot filling factor $\nu_\mathrm{AD} = 2$, in which antidot transport experiments are well-described by a non-interacting model of single-particle antidot states \cite{mace1995general}. The parameter $\nu_\mathrm{AD}$ defines the number of spin-polarized edge modes that circle the antidot. It is typically set by voltages applied to nearby gates to form constrictions where the filling factor is lower than in the bulk; see Fig~\ref{fig:ADdevice}B. When $\nu_\mathrm{AD} = 2$, the antidot is equivalent to a dot of holes in the lowest Landau level of the 2DES, where a ``ladder'' of discrete orbital states exists for each of two spin configurations. The orbital, single-particle energy spacing, $\Delta E_\mathrm{SP}$, is determined by the antidot's electrostatic profile and magnetic confinement, and the two ladders are separated by the Zeeman energy; see Fig.~\ref{fig:ADPhys}A.

Viewed as dots of holes, the quantized energy spectra of quantum antidots are analogous to the spectra of similarly sized quantum dots \cite{Reimann2002qdreview}. However it is known that in quantum dots there is a maximum particle number, $N_\mathrm{max} \approx 40$, beyond which transitions occur in the dots' core through the population of higher Landau levels, rather than at the edge \cite{Oosterkamp1999qdmdd,Ciorga2002qdspinpolarization}. For antidots, in contrast, there are no ``higher'' Landau levels available for holes in the core, and the $\nu_\mathrm{AD} = 2$ configuration is preserved to arbitrary $N$. The ground-state configuration, in which all hole states are filled up to the maximum orbital state defined by the Fermi level, is known as a maximum-density droplet \cite{MacDonald1993mdd}. Notably, this ground state remains an exact eigenstate of the many-body Hamiltonian even in the presence of electron-electron interactions.

We model the electrostatics of a quantum antidot as an equivalent capacitor network; see Fig.~\ref{fig:ADdevice}C. Changing the gate voltage $V_G$ shifts the antidot energy levels up or down as a whole, but generally does not change their spacing. In this way, sweeping through the gate voltage is analogous to moving through the periodic table for natural atoms by increasing the nuclear charge \cite{kastner1993artificial}. Tuning the bias between source $V_S$ and drain $V_D$ voltages changes the relative energy levels of the electrons in the metallic leads, and also shifts the antidot states through capacitive interactions. By convention, we assume that the source contact is connected to ground, and that a bias is applied to the drain.

In principle, transport can occur \textit{via} tunneling between the antidot states and any of the extended edge modes ($P$, $P^\prime$, $Q$, and $Q^\prime$ in Fig.~\ref{fig:ADdevice}B). The tunneling rate across the constrictions (to $P$ and $P^\prime$) can be precisely controlled using gate voltages that adjust the constriction widths. Here, we assume that tunneling to $P$ and $P^\prime$ is fully suppressed, such that transport only results from tunneling to the higher Landau levels $Q$ and $Q^\prime$. In experiments, such tunneling events are detected as an increase in current flowing across the device. A further advantage of studying quantum antidots in this configuration is that the tunneling rates are comparable for both spin species, whereas in quantum dots usually the transport is dominated by a single spin species due to the topology of edge channels.

For transport to occur, there must be either an occupied state in the source and an unoccupied state in the drain, or an occupied state in the drain and an unoccupied state in the source. Furthermore, the energy difference of a transition between antidot states, $\mu_\text{antidot}(N)=E(N) - E(N-1)$, must match the window defined by the relative chemical potentials of the source, $\mu_{S}$, and drain, $\mu_{D}$. At zero bias, where $\mu_{S} = \mu_{D} = 0$, transport occurs only for $\mu_\text{antidot}(N) \approx 0$. Otherwise, transport is blocked by the Coulomb blockade. When $\mu_{S}$ or $\mu_{D}$ is varied, transport occurs when $\mu_\text{antidot}$ is within the transport window defined by $\mu_{S}$ and $\mu_{D}$. These features give rise to a characteristic pattern of Coulomb diamonds in the differential conductance as a function of antidot gate voltage (which adjusts $\mu_\text{antidot}$) and source-drain bias (which adjusts $\mu_D$); see Fig.~\ref{fig:ADPhys}B-C.

The energy separating two transitions with different $N$ is given by $\Delta E = \mu_\text{antidot}(N+1) - \mu_\text{antidot}(N) = E_C + \Delta E_{ex}$, where $E_C=e^2/C$ is the electrostatic charging energy and $\Delta E_{ex}$ is the quantum mechanical contribution from the single-particle energy levels. $\Delta E$ can be directly measured from the Coulomb diamond pattern by identifying the drain voltage at the intersection point ``D'' through $\Delta E = -e \Delta V_D$, and it is related to the width of the Coulomb diamonds in gate voltage through the capacitive factor, $\Delta V_G = \frac{C_G}{C}\Delta V_D$. Additionally, transitions involving excited electronic states can contribute to transport \cite{Michael2006}; these transitions are observed as lines parallel to the Coulomb diamonds as indicated by point ``C'' in Fig.~\ref{fig:ADPhys}B.

\subsection{\label{sec:level2:PhysFea} Physical features of interest}

We consider two physical features of the system that we hypothesize are sensitive to the topological architecture of the energy landscape of antidot state transitions: the current, $I$, and the differential conductance, $G=\frac{dI}{dV_D}$. The current is non-zero when the number of available transitions in the antidot between $\mu_{S}$ and $\mu_{D}$ is non-zero (see Fig.~\ref{fig:ADPhys}C), so studying current indicates the accessibility of energy states. Meanwhile, conductance depends on the alignment of various state transitions with $\mu_{S}$ and $\mu_{D}$, and its value (either positive or negative) reflects changes in the transport configuration.

In some situations, spin-dependent selection rules result in negative differential conductance, where the current flowing through the device drops as the bias is increased. Negative differential conductance is a signature of spin blockade, where the energetic availability of a new tunneling transition causes a change in the antidot's steady-state spin configuration, which subsequently blocks the dominant transport channel and causes the current to drop \cite{weinmann1995spin,bassett2009probing}. (Note the distinction: in regions that are Coulomb blockaded, both current and conductance are low, whereas in regions that are spin blockaded, there exists a measurable current but a negative conductance.) 

Spin blockade is a form of frustration in quantum transport, where competing transport channels cause nearly-degenerate quantum configurations to shift or become unstable. Thus, measurements of current and conductance give us insight into whether transport occurs and whether the system is frustrated. By applying techniques from network science, we seek to relate the topological properties of the energy state networks to these physical measurements of transport.

\subsection{\label{sec:level2:CompMod} Computational model of quantum transport through an antidot}

\indent To study the transport properties of a quantum antidot weakly coupled to two metallic leads by spin-dependent tunneling barriers, we model transitions between different occupation states of the antidot as a Markov chain. A master equation describes the stochastic evolution of the system, and the solution to the master equation provides the steady-state probability occupations for the antidot's noninteracting quantum configurations, $\lvert s\rangle =
\lvert \lbrace n_{\ell,\sigma}\rbrace\rangle$. The configurations are labeled by particle occupation numbers $n_{\ell,\sigma}=0,1$ for the single-particle state with orbital and spin quantum numbers $\ell$ and $\sigma$.

\indent The transition rates $\gamma_{s' \rightarrow s}$ from antidot configuration $s'$ to $s$ are calculated according to a combination of antidot selection rules and Fermi's golden rule; see Sec. IA in the Supplementary Materials for a full derivation of the transition rates. In principle, the number of possible states for an antidot with $N$ holes and $M$ available orbital levels, $\binom{2M}{N}$, is exceedingly large. However, many configurations can be ignored since they are not energetically accessible, and many transition rates are zero due to selection rules. In Sec.~IC of the Supplementary Materials, we describe a method to enumerate the energetically accessible configurations according to their particle number, $N$, and their total spin projection, $S_z$. The resulting transition rate matrix, $\mathbf{R}$, is defined by $R_{ij} = \gamma_{s_is_j}=\gamma_{s_j \rightarrow s_i}$, where $i$ and $j$ represent different configurations; see Fig.~\ref{fig:NetRep}A. We seek the equilibrium configuration where the total transition rate into and out of each state is equal,
\begin{equation}
0 = \sum_{s}[\gamma_{ss'}P(s') - \gamma_{s's}P(s)]~,
\label{eq:GammaConserv}
\end{equation}
where $P(s)$ represents the equilibrium occupation probability of state $s$. Combining Eq.~\ref{eq:GammaConserv} with the normalization condition $\sum_{i}^{m}P(s_i) = 1$ for $m$ available states, we obtain the master equation in matrix form, 
\begin{equation}
\begin{pmatrix}
  \sum_{i}\gamma_{s_{i}s_{1}} & -\gamma_{s_{1}s_{2}} & \cdots & -\gamma_{s_{1}s_{m}} \\
-\gamma_{s_{2}s_{1}} & \sum_{i}\gamma_{s_{i}s_{2}} & \cdots & -\gamma_{s_{2}s_{m}} \\
  \vdots  & \vdots  & \ddots & \vdots  \\
  -\gamma_{s_{m}s_{1}} & -\gamma_{s_{m}s_{2}} & \cdots & \sum_{i}\gamma_{s_{i}s_{m}} \\
  1 & 1 & \cdots & 1
 \end{pmatrix}
 \begin{pmatrix}
 P(s_{1}) \\
 P(s_{2}) \\
 \vdots \\
 P(s_{m})
 \end{pmatrix} = 
 \begin{pmatrix}
 0 \\
 0 \\
 \vdots \\
 0 \\
 1
 \end{pmatrix}.
 \label{eq:MasterEq}
\end{equation}
The matrix on the left of Eq.~\ref{eq:MasterEq} is obtained by adding diagonal elements to the rate matrix $\begin{bf}R\end{bf}$ to impose conservation of rates as well as an additional row of ones to enforce normalization. The solution to Eq.~\ref{eq:MasterEq} yields the equilibrium occupation probabilities $\begin{bf}P\end{bf}$. From $\begin{bf}P\end{bf}$, we can compute the current flowing from each spin-polarized reservoir, and subsequently the spin-resolved conductance at finite bias; see Sec. IB-C in the Supplementary Materials for a derivation of the expressions used to compute spin-resolved $I$ and $G$. 

\indent Using this computational model, we can input a series of experimental parameter settings including gate voltages, drain-source bias, magnetic field, and temperature, to predict the current and conductance. The settings chosen in this work are motivated by spin-resolved transport experiments in which the underlying physical parameters (\textit{e.g.}, $\Delta E_\mathrm{SP}$, $E_\mathrm{Z}$, and the spin-dependent tunneling rates) have been well characterized \cite{bassett2009probing}. Unless indicated otherwise, the temperature is \SI{50}{\milli\kelvin}, $\Delta E_\mathrm{SP}=\SI{30.7}{\micro\electronvolt}$, $E_Z=\SI{45.8}{\micro\electronvolt}$, and $E_C=\SI{85}{\micro\electronvolt}$.

We calculate maps of $I$ and $G$ as a function of $V_{DS}$ and $V_G$, typically with 24 settings of $V_{DS}$ and 72 settings of $V_G$ to span a pair of Coulomb diamonds. In the noninteracting picture, subsequent transitions involve the tunneling of opposite spin species, due to the interleaved energy-ladder diagram in Fig.~\ref{fig:ADPhys}A. This particular antidot configuration also exhibits frustration due to spin-blockade at certain voltage settings, evidenced by regions of negative differential conductance. Therefore we can analyze the energy state networks in such a way as to distinguish between frustrated and non-frustrated regimes.

\subsection{\label{sec:level2:NetRep} Network representation}
For each voltage setting, we model the energy landscape as electrons tunnel into and out of the antidot as a network. We are primarily interested in networks in which transport occurs, so we exclude networks corresponding to voltage settings in which there is no measured current (\textit{i.e.}, inside the Coulomb diamonds). In our voltage space, we construct 1271 networks corresponding to different combinations of drain-source and gate voltage settings.

\begin{figure}
  \centering
  \includegraphics[width=8.6cm]{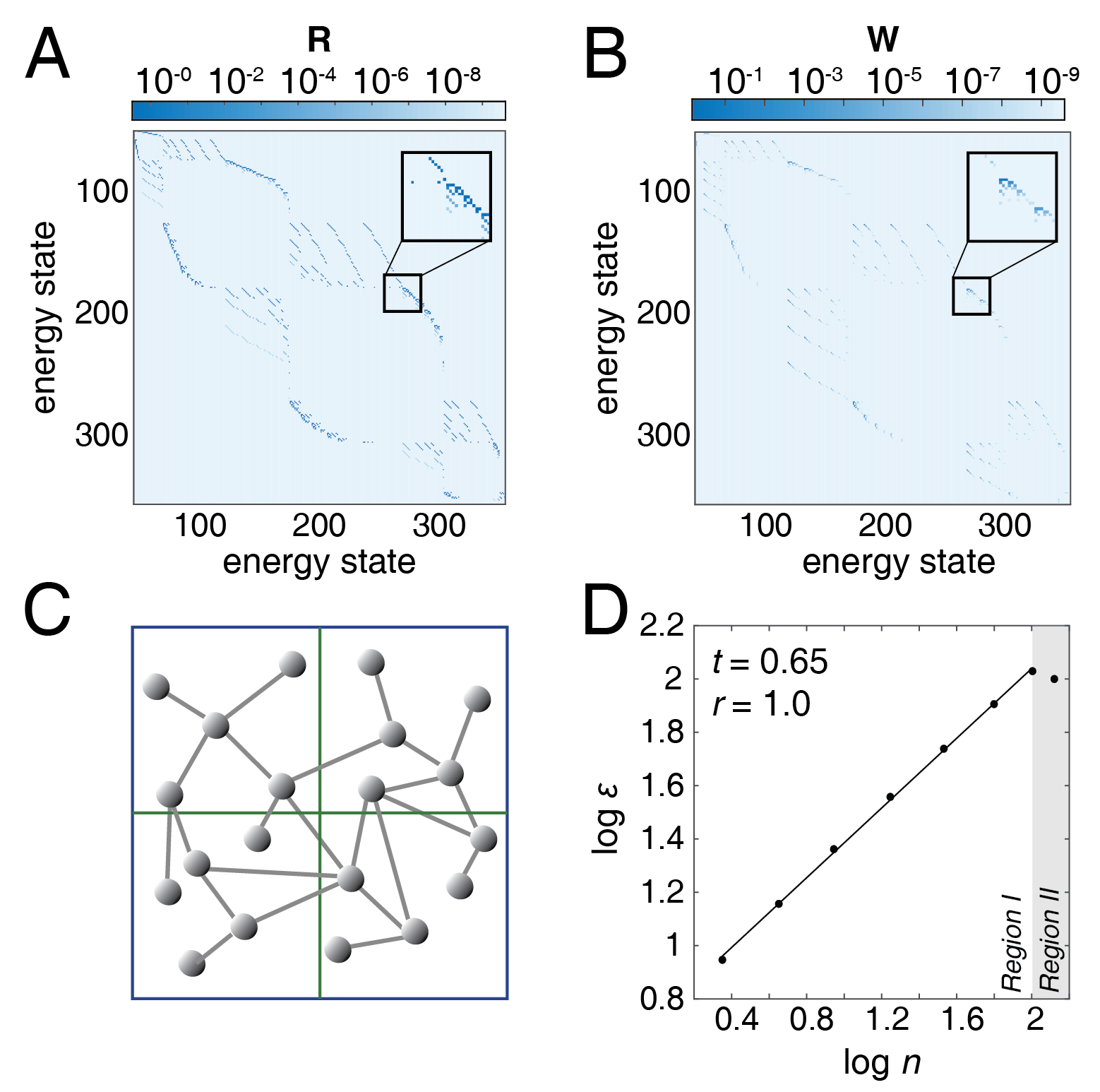}
  \caption{\textbf{Topological Rentian scaling.} \textbf{A.} Example transition rate matrix \textbf{R} determined by antidot selection rules and Fermi's golden rule. \textbf{B.} Weight matrix \textbf{W} corresponding to the transition rate matrix \textbf{R} displayed in panel \textbf{A}. The weight matrix \textbf{W} is determined by weighting \textbf{R} using its probability vector \textbf{P} as described in the text, thresholded to exclude numerical inaccuracies in values near zero. In both panel \textbf{A.} and panel \textbf{B.}, lone nodes were excluded for visualization. We perform our network analyses on a binarized version of the weight matrix \textbf{W}. \textbf{C.} Schematic of the topological partitioning process where the network is recursively partitioned into halves, quarters, and so on with an algorithm (\textit{hMETIS}) that minimizes the number of edges crossing each partition \cite{karypis2000multilevel}. With each iteration of the partitioning algorithm, the number of nodes within a partition ($n$) and the number of edges ($\varepsilon$) crossing the boundary of each partition are recorded. A linear relationship between the number of nodes and the number of edges in log-log space indicates the presence of topological Rentian scaling. \textbf{D.} Example of the topological Rentian scaling relationship for a single transport matrix of an antidot.
  }
  \label{fig:NetRep}
\end{figure}

\indent In each network, nodes represent the possible energy states that the system can occupy, and edges represent the possible transitions between energy states. We first threshold the probability vector $\mathbf{P}$ in order to remove numerical inaccuracies in probability values close to zero, while retaining a fully-connected network (see Sec. IIA and Fig. S2 in the Supplementary Materials). We form a weight matrix, $\mathbf{W}$, from $\mathbf{R}$ and $\mathbf{P}$ such that $W_{ij}= R_{ij}P_{j} + R_{ji}P_{i}$; see Fig.~\ref{fig:NetRep}B. The weight matrix is non-directional ($W_{ji} = W_{ij}$); each element $W_{ij}$ comprises the sum of the forwards and backwards transition rates between states $i$ and $j$, respectively weighted by the occupation probability of the original state. This formulation decreases (increases) the relative weight of transitions between energy states with low (high) occupation probabilities. Then, we binarize $\mathbf{W}$, which yields an unweighted adjacency matrix, $\mathbf{A}$, where $A_{ij} = 1$ if there is an edge between nodes $i$ and $j$, and $A_{ij} = 0$ otherwise. We remove any nodes that are not connected to any other node by an edge. We define the size of this network to be the number of nodes remaining after the lone nodes have been removed (See Fig.~\ref{fig:RentScal}B), and we perform our network analyses on these reduced \textbf{A} matrices.

\subsection{\label{sec:level2:TopRent} Topological Rentian scaling}
\indent Rent's rule is an empirical power law first discovered in integrated circuit designs that describes a scaling relationship between the number of external signal connections (edges) in a block of a computer logic graph and the number of components (nodes) in the block \cite{christie2000interpretation}. Despite these rather specific origins, fractal Rentian scaling can be examined as a general organizational principle in complex networks. Rentian scaling has previously been demonstrated to be exhibited by the London Underground \cite{sperry2016rentian}, neuronal networks \cite{bassett2010efficient}, and vasculature systems \cite{papadopoulos2018comparing} in both topological and physical space. In our analysis, we focus exclusively on topological Rentian scaling, as our networks\,---\,which represent transitions between accessible energy states\,---\,have no spatial arrangement. Although our networks lack a spatial embedding, the topology is constrained by physical laws, including the conservation of energy, spin, and angular momentum.

\indent Formally, the power law for topological Rentian scaling can be expressed as 
\begin{equation}
\varepsilon = kn^{t}
\end{equation}
\noindent where $\varepsilon$ is the number of edges that cross a topological partition of the network (see Fig.~\ref{fig:NetRep}C), $n$ is the number of nodes in the partition, $0 \leq t \leq 1$ is the Rent's topological exponent, and $k$ is the Rent coefficient. A higher Rent's topological exponent indicates a higher dimensionality of the interconnect topology \cite{Ozaktas:1992a,Stroobandt:2011a}. 

\indent To determine whether the quantum transport networks exhibit Rentian scaling, we used the \textit{hMETIS} software \cite{karypis2000multilevel} to recursively partition each network in topological space. At each level of partitioning, we recorded the number of nodes $n$ in each partition and the number of edges $\varepsilon$ that intersect the partition boundaries; see Fig.~\ref{fig:NetRep}C. When estimating the Rent's exponent, it is important to note that the partitioning analysis typically results in two qualitatively distinct regions. The existence of the power law has been empirically confirmed in a region designated \textit{Region I}, where $n$ is much smaller than the size of the network. However, Rent's rule overestimates the interconnection complexity in a region designated \textit{Region II}, where the number of partitions is small, or $n$ is large \cite{landman1971pin}. 

\indent Because we are interested only in \begin{it}Region I\end{it}, we follow prior work by excluding points where $n$ is greater than half of the total number of nodes in the network \cite{bassett2010efficient}. We estimated the slope of the $n$ versus $\varepsilon$ values in log-log space using a linear regression. To determine whether the networks exhibit topological Rentian scaling, or in other words, whether the relationship between $n$ and $\varepsilon$ is linear in log-log space, we compute the correlation between $n$ and $\varepsilon$ as a second metric quantifying the goodness of fit of the linear model to the data; see Fig.~\ref{fig:NetRep}D. For more information on how we estimated the values of Rent's topological exponent, see Sec. IIB and Fig. S3 in the Supplement.

\section{\label{sec:level1:Results} Network topology captures physical features of quantum transport}

We explored the relationships between the topological properties of quantum transport networks\,---\,specifically the existence of Rentian scaling and Rent's exponent\,---\,and the system's physical characteristics as manifested by current and conductance.

\begin{figure*}
  \centering
  \includegraphics[width=\textwidth]{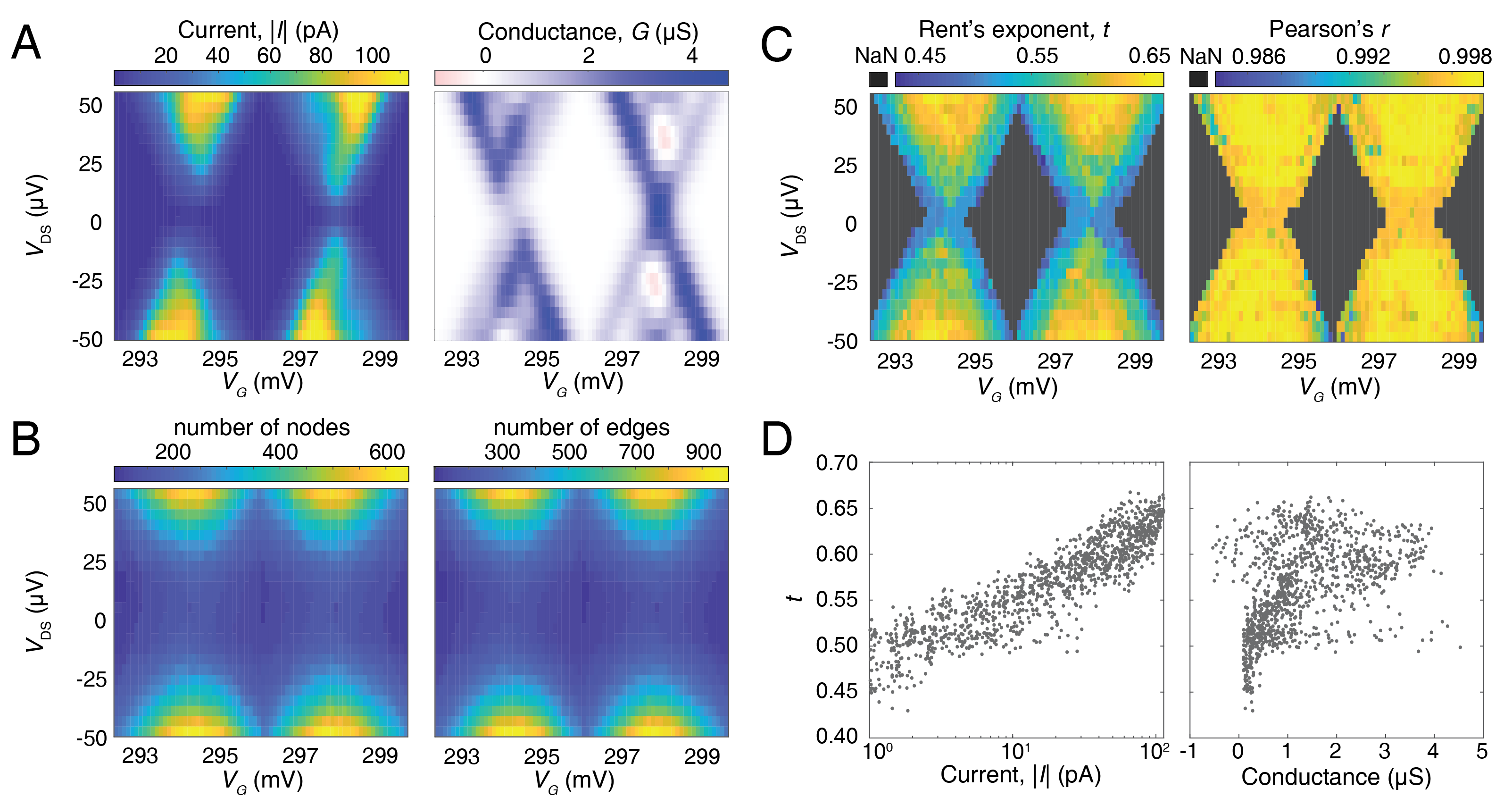}
  \caption{\textbf{Topological Rentian scaling in antidot transport matrices.} \textbf{A.} Quantum transport calculations. The absolute value of current (\emph{Left}) and conductance (\emph{Right}) are calculated as a function of drain voltage ($V_D$) and gate voltage ($V_G$). Both current and conductance are zero inside the Coulomb diamonds where transport is prevented by Coulomb blockade. Regions where the current is nonzero and the conductance is negative (red in the right-hand plot) reveal frustration due to spin-blockade effects. \textbf{B.} Network size. The number of nodes (\emph{Left}) and the number of edges (\emph{Right}) for networks corresponding to the transport configurations in panel \textbf{A}. \textbf{C.} Topological complexity. Rent's topological exponent (\emph{Left}) and the corresponding Pearson correlation coefficient (\emph{Right}) between the number of nodes and edges in log-log space are shown for networks corresponding to the transport configurations in panel \textbf{A}. Since we exclude networks corresponding to voltage settings under which the antidot experiences a current $< 1$ pA, $t$- and $r$-values for the excluded networks are displayed as NaN. \textbf{D.} Scaling relationships. Rent's topological exponent is plotted versus the absolute value of the current (\emph{Left}) and the conductance (\emph{Right}).}
  \label{fig:RentScal}
\end{figure*}

\subsection{\label{sec:level2:rent} Topological complexity of the antidot's energy landscape}

\indent As described in Sec.~\ref{sec:level1Methods}, we performed quantum transport calculations and constructed energy-state network representations for a range of voltage configurations spanning a full cycle of two Coulomb diamonds; see Fig.~\ref{fig:RentScal}A. The differences between adjacent Coulomb-blockade patterns arises from the antidot's alternating spin configuration as successive states in the single-particle ladder are filled; see Fig.~\ref{fig:ADPhys}A. 

For each voltage setting, we examined the Rentian scaling behavior and Rent's exponent. All networks corresponding to voltage settings under which the antidot experienced a current greater than 1 pA exhibit topological Rentian scaling. The value of Rent's topological exponent $p_t$ ranges from $0.4479$ to $0.6745$, and the Pearson correlation coefficient ranges from $0.9878$ to $0.9999$; see Fig.~\ref{fig:RentScal}A \& C. When reporting an exponent for the Rentian scaling relation, it is important to assess the degree to which a power law fit is appropriate. We observed that power laws provided better fits to the topological partition data than comparable exponential, logarithmic, linear, quadratic, or cubic models; see Sec. IIIA and Table 1 in the Supplementary Materials.

\indent From a physical perspective, a key finding of our analysis is that the topological exponent scales with the log of the current ($r = 0.8914, p < 0.005$); see Fig.~\ref{fig:RentScal}D, left. Thus, energy state transition networks with higher topological complexities tend to exhibit higher currents. Furthermore, Rent's topological exponent generally increases as conductance increases, although the scatter is larger; see Fig.~\ref{fig:RentScal}D, right. Broadly, the observed relationship between the topological network property of Rentian scaling and physical measurements of transport suggest that the network approach captures physically relevant information about the quantum system. Outliers to this overall trend\,---\,especially networks that exhibit large values for Rent's exponent when the current or conductance is low\,---\,are the subject of further exploration. Notably, we did not observe strong relationships between the physical quantities of current and conductance and other network measures, including network size, density, topological efficiency, average clustering coefficient, and assortativity; see Sec. III B in the Supplementary Materials.

\subsection{\label{sec:level2:frust} Topological complexity marks frustration}

%\remove[LP]{\indent We observed that networks corresponding to voltage settings under which the antidot experiences low levels of conductance (less than $0.5 \mu S$) appear to separate into two groups naturally: one with lower values of Rent's exponent ($t < 0.56$) and one with higher values of Rent's exponent ($t > 0.57$); see Fig.~}\ref{fig:FrustTemp}.\remove[LP]{The latter group notably includes all configurations where the conductance of either the spin-up or spin-down electrons is negative.}\note[LP]{I think this paragraph and values of the Rent exponents here are referring to an older version of the manuscript?}

As described in Sec.~\ref{sec:level2:PhysFea}, the antidot model exhibits a form of frustration known as spin blockade, where the availability of additional energy states due to an increase in source-drain bias leads to a situation where current is blocked due to the Pauli exclusion principle. This effect causes the current to drop, resulting in negative differential conductance. In this section, we examine the relationships between the topological complexity of the quantum transport networks and the physical phenomenon of frustration.

\begin{figure*}
  \centering
  \includegraphics[width=\textwidth]{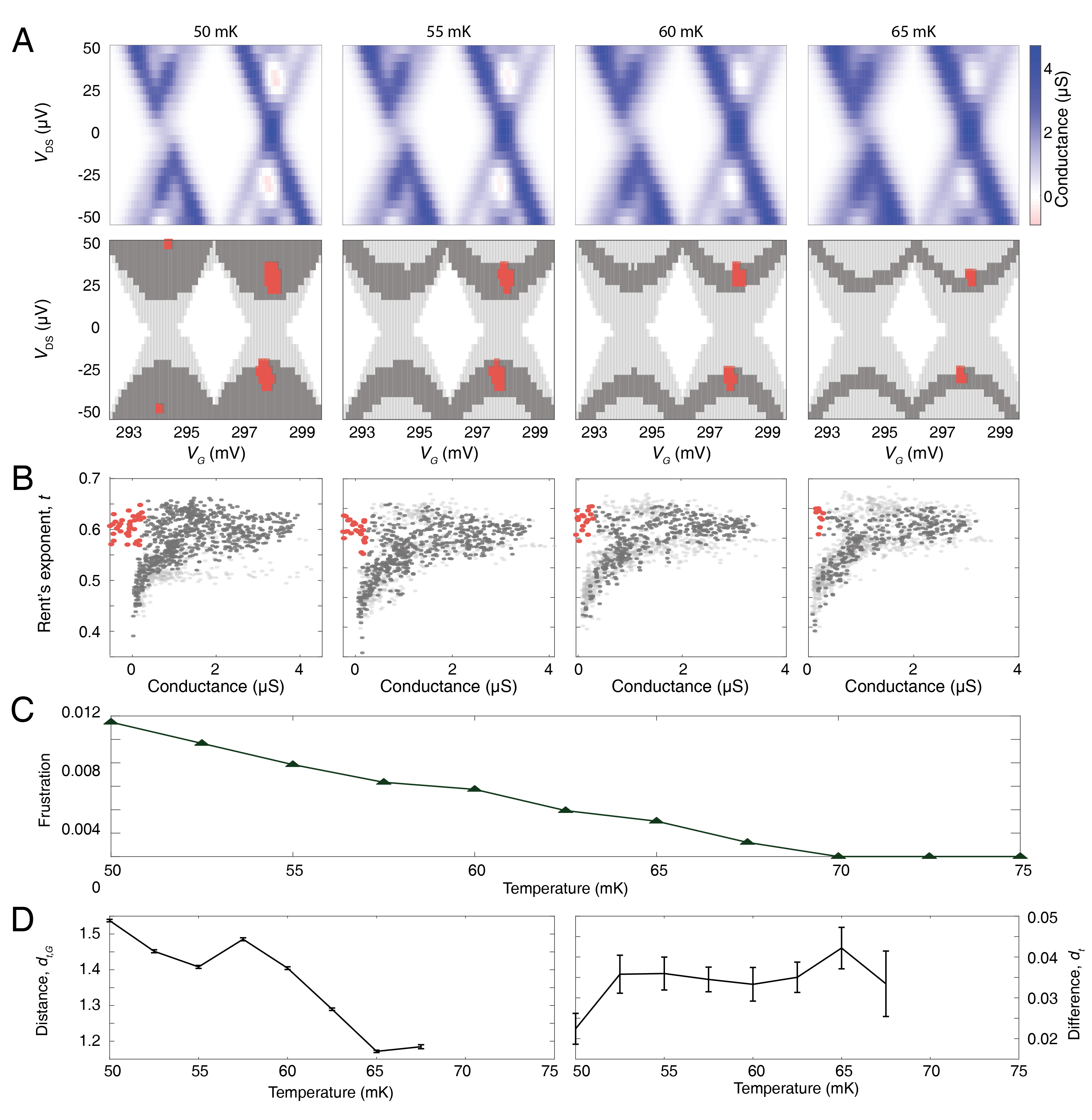}
  \caption{\textbf{Relationship between frustration and topological complexity}. \textbf{A.} \textit{(Top)} Total conductance simulated for temperature settings at 50 mK, 55 mK, 60 mK, and 65 mK. \textit{(Bottom)} Frustrated (red) and corresponding control groups (dark gray) highlighted in the $V_D$ versus $V_G$ space corresponding to the plot above. Voltage settings corresponding to antidots that experience a measurable current but result in networks with sizes outside the range of the frustrated networks are shown in light gray. \textbf{B.} Rent's topological exponent as a function of conductance for the set of networks from the corresponding simulation in panel \textbf{A}, with the frustrated and non-frustrated control groups highlighted in red and dark gray, respectively. \textbf{C.} Frustration, measured as the fraction of voltage settings in a given simulation under which the spin-up or spin-down conductance is negative, as a function of temperature. \textbf{D.} The distance, $d_{t,G}$ \textit{(Left)}, in the space of conductance versus Rent's topological exponent between frustrated and non-frustrated control networks, as a function of temperature, and the difference, $d_{t}$ \textit{(Right)}, of the average Rent's exponent between the frustrated and non-frustrated groups. Based on $p$-values from non-parametric permutation tests, $d_{t,G}$ and $d_t$ are statistically significant for all temperatures (see Sec. IIC in the Supplmentary Materials for details). For temperatures $\geq 70$ mK, $d_{t}$ and $d_{t,G}$ are not defined, as there is no longer a frustrated group of networks.
  }
  \label{fig:FrustTemp}
\end{figure*}

Our computational quantum transport model separately predicts the current and conductance due to tunneling by spin-up and spin-down electrons; see Fig.~S1 in the Supplementary Materials. In some cases, the transport through one spin channel can exhibit frustration (\textit{i.e.}, negative conductance) even when the total sum of the conductance across both spin-up and spin-down channels is positive. In order to capture these subtle effects, we first defined frustrated configurations as those voltage settings where either of the spin-resolved differential conductance values were negative. The top row of Fig.~\ref{fig:FrustTemp}A shows the total conductance as a function $V_{DS}$ and $V_{G}$ for several different temperature settings, and in the bottom row of Fig.~\ref{fig:FrustTemp}A, frustrated configurations are highlighted as red points. 
%\remove[LP]{These configurations are highlighted as red points in Fig.~}\lp{\ref{fig:FrustTemp}}\remove[LP]{A-B. Clearly, the set of networks with low conductance and high values of Rent's exponent corresponds to these frustrated voltage settings.}

To study the relationship between network structure -- as quantified by topological complexity -- and frustration, we plot all of the networks from Fig.~\ref{fig:FrustTemp}A in the 2-dimensional space of Rent's topological exponent \emph{vs.} conductance (see Fig.~\ref{fig:FrustTemp}B). The set of networks corresponding to frustrated voltage settings are again highlighted in red. Focusing first on a single value of temperature where frustration is markedly present in the system (for example, at 50 mK), we see clearly that frustrated configurations have low values of conductance (by definition), but interestingly, also have consistently high values of the Rent's exponent. This observation suggests that the degree of topological complexity in the quantum transport networks may be a marker of frustration in the underlying system.

To more rigorously test the hypothesis that topological network complexity correlates with frustration in quantum transport, we performed a nonparametric permutation test to quantify the statistical significance of the difference, $d_{t}$, between the average values of the Rent's topological exponent for the frustrated and non-frustrated configurations. Note that we define $d_{t}$ such that positive values imply that the average Rent's exponent is greater for the set of frustrated networks relative to the non-frustrated control set. The group of frustrated networks is determined as described above based on the observation of negative differential conductance, and we further identify a comparable set of non-frustrated networks to use as a control group. The control group consists of networks having the same range of sizes (number of nodes) as the set of frustrated networks. Hence, the frustrated and non-frustrated control group of networks are characterized by a similar number of available energy states, but the pattern of state transitions, or the distribution of edges throughout the network, is different. The control groups of non-frustrated networks are highlighted in the bottom row of Fig.~\ref{fig:FrustTemp}A \& in Fig.~\ref{fig:FrustTemp}B in dark gray.
% \note[LB]{Would it be possible to highlight this control group of non-frustrated networks in Fig 5 with a different color?}

In comparing the set of frustrated networks to the non-frustrated control group in Fig.~\ref{fig:FrustTemp}B, we observe that the frustrated configurations are always localized in a region of high topological complexity, while the control group exhibits a large spread along the Rent's-exponent axis. We indeed find that the difference $d_{t} >0$ and that it is a statistically significant quantity to distinguish between the two sets of networks shown in Fig.~\ref{fig:FrustTemp}B ($d_t = 0.022$, $p = 1.000\times10^{-3}$ at $50$mK), suggesting that certain architectural features of the state transition networks are sensitive to and can distinguish between different physical aspects of transport. For a more detailed exposition on our approach to calculate $d_{t}$ and $p$, see Sec. IIC in the Supplementary Materials.

\subsection{\label{sec:level2:temp}Relationships between temperature, frustration, \& topological complexity}

\indent To further examine the physical significance of the relationship between frustration and energy state transition networks with an enhanced topological complexity, we ran our model of sequential quantum transport over the same voltage space at different temperatures; see Fig.~\ref{fig:FrustTemp}A. The antidot configurations are exactly the same in each simulation; the temperature setting determines the distribution of filled and empty states in the metallic leads, and therefore the number of antidot energy states that are accessible for tunneling transitions. The effect of increasing temperature is to blur out fine structure due to tunneling through excited states and, eventually, to overcome spin blockade by providing additional tunneling pathways.

We quantified the amount of frustration present in a given simulation as the fractional number of frustrated networks, obtained by dividing the number of voltage settings under which either spin-up or spin-down conductance is negative by the total number of voltage settings. Fig.~\ref{fig:FrustTemp}C shows that the fractional amount of frustration decreases as the temperature is increased, eventually reaching a situation for temperatures $\geq\SI{70}{\milli\kelvin}$ where no negative differential conductance is observed. However, even though the overall amount of frustration decreases with temperature, we still observe a clear separation between the frustrated and non-frustrated configurations with increasing temperature when the networks are visualized in the 2-dimensional space of Rent's exponent \emph{vs.} conductance; see Fig.~\ref{fig:FrustTemp}B.

Using temperature to modulate the spin-blockade mechanism by which the antidot system is frustrated, we thus sought to further investigate the relationships between temperature, frustration, and topological complexity. We began by examining the temperature dependence of the distance between frustrated and non-frustrated networks as measured in the two-dimensional space spanned by Rent's exponent \emph{and} conductance, $d_{t,G}$. We observe that $d_{t,G}$ decreases as the temperature increases (see the left panel of Fig.~\ref{fig:FrustTemp}D), but that the distance between the two groups remains statistically significant for temperatures up to 67.5mK (see Sec. IIC in the Supplementary Materials for details on how statistical significance was determined). For temperatures $\geq$70~mK, where our measure of frustration vanishes, $d_{t,G}$ is no longer defined, as there are no frustrated networks from which to measure a distance.

To better understand the decreasing trend of $d_{t,G}$ with temperature and to isolate how the difference in topological complexity between the frustrated and non-frustrated networks behaves as a function of temperature, we next consider the difference $d_{t}$ of the average Rent's exponent between the frustrated and non-frustrated configurations. The right panel of Fig.\ref{fig:FrustTemp}D shows $d_{t}$ \emph{vs.} temperature. Interestingly, we find that the difference $d_{t}$ remains significantly greater than zero for all temperatures (up to 70mK, at which point frustration disappears), implying that the average Rent's exponent of the frustrated networks remains higher than that of the control set across temperatures. See Sec. IIC in the Supplementary Materials for a detailed description of how statistical signifcance was assessed. In other words, Rent's exponent\,---\,which is determined solely from the topological organization of the state transition networks\,---\,continues to capture aspects of the system associated with frustration, even as the phenomenon of frustration itself is mitigated by additional tunneling pathways that arise with increasing temperature.

Our findings demonstrate that the topological complexity of the quantum transport network is not determined solely by the quantum-mechanical configurations of the antidot (or, more generally, the configurations of any mesoscopic quantum system), but rather by the available transitions between these states as determined by energy and spin selection rules. Moreover, the results suggest a link between network complexity as measured by Rent's exponent and frustration as exhibited in spin blockade. The pattern of available transitions between energy states in the frustrated networks appears to be driving the increased topological complexity of the network compared with the non-frustrated networks in our control group. Since the control group is the set of non-frustrated networks with access to a number of energy states that is comparable to that of the frustrated networks, and the number of edges is highly correlated with the number of nodes (see Supplementary Figure S5), the difference in topological complexity between the two groups must be driven by qualitative differences in the distribution of edges throughout the network rather than by differences in the number of edges or nodes between the two groups. 

\begin{comment}
the existence (or non existence) of system frustration is a sufficient explanation for the high (or low) complexity of the quantum transport network.
\end{comment}

\begin{comment}
\begin{figure}
  \centering
  \includegraphics[width=0.45\textwidth]{Fig12v2.png}
  \caption{\textbf{Relationship between frustration and topological complexity}. \textbf{A.} Total conductance, simulated for temperature settings at 50 mK, 57.5 mK, and 72.5 mK. \textbf{B.} Frustration, measured as the number of voltage settings under which the spin-up or spin-down conductance is negative, normalized by the total number of voltage settings, as a function of temperature. \textbf{C.} The distance, $d$, in the space of conductance versus Rent's topological exponent between frustrated and non-frustrated networks that experience comparable levels of current, as a function of temperature. For all values of $d$, we found $p = 0$ by non-parametric permutation test. For temperatures $\geq 70$ mK, $d$ is not defined because there is no frustrated group of networks from which to measure a distance to a control group of networks.
  }
  \label{fig:Temp}
\end{figure}
\end{comment}

\section{\label{sec:level1:Discussion} Discussion}

\subsection{The physics of many-body quantum systems}

Many-body quantum systems are difficult to simulate exactly due to the exponential scaling of complex parameters with the number of particles in the system. For instance, considering only the spin states of $N$ particles, the size of the relevant Hilbert space is $2^N$, and the space is even bigger if the orbital states are also included. The antidot system considered here features $N\approx130$ holes, and keeping track of so many parameters for exact quantum-mechanical calculations is intractable. Instead, we use effective models to describe collective excitations of the quantum system with fewer degrees of freedom \cite{MacDonald1993mdd,Reimann2002qdreview,bassett2009probing}.
For example, the semiclassical single-particle model considered here, together with physical limitations due to energy and spin conservation, reduces the number of antidot configurations to a more computationally tractable set of $\approx$3400 states. Subsequent reductions based on a threshold on $\mathbf{P}$ yields $\approx$200-700 states, depending on the voltage setting (see Sec.~IIA in the Supplementary Materials). 

Nonetheless, even these simplified effective theories are too complex to treat analytically, and whereas computational transport models can simulate experimental measurement parameters such as current and conductance, there is a paucity of computational methods to capture or predict \emph{qualitative} features of the physics such as frustration, which is a unique dynamical phenomenon involving a highly degenerate energy landscape. In this work, we demonstrate how a network science approach can identify parameter settings that exhibit frustration in quantum transport. Besides providing a new computational tool that is tractable for such investigations, our analysis of topological complexity also illuminates how the structure of state transitions contributes to the dynamics of many-body quantum systems and how network science can be used to model their behavior.

\subsection{The network architecture of quantum transport networks}

\indent The quantum antidot is a transport system governed by physical constraints in the form of conservation laws. Its topological structure can be parsimoniously represented by a network model in which energy states are interconnected by allowable transitions. The constrained nature of the energy state transition network of an antidot makes it a natural one in which to examine Rentian scaling, a common organizational principle shared by other kinds of efficient physical and biological communication and transport systems shaped by economic and evolutionary constraints. Notably, however, the quantum transport networks, unlike many of the networks that have previously been demonstrated to exhibit Rentian scaling, do not exist in physical space. 

\indent While in some systems the presence of Rentian scaling is indicative of evolutionary and economic constraints converging toward a common scheme of optimization, here Rentian scaling is indicative of the statistical homogeneity arising from conservation laws. Generally in studies of this type, the term statistical homogeneity is used to imply that topological quantities such as the average number of edges per node are independent of the node's physical position within the circuit \cite{christie2000interpretation}. In our particular case, we use the term to indicate homogeneity of topological properties across scales, where here the term ``scale'' refers to separation in energy/spin configurations rather than in physical space. Notably, prior theoretical work agnostic to the domain of application has demonstrated that statistical homogeneity is a sufficient condition to derive the power-law relationship known as Rent's law from first principles \cite{christie2000interpretation, greenfield2010communication}. 

Our observation that the energy-state transition landscape of quantum systems is fractal in nature has important implications for accessible and efficient strategies for network-based control of the quantum dynamics. Recent studies have demonstrated that the self-similarity inherent in fractal networks is beneficial in obtaining exact analytical solutions for specific control strategies and for identifying the optimal set of driver nodes supporting those strategies \cite{liu2011controllability, li2014controllability}. Future efforts expanding our network modeling approach to assess the control of the energy state transition networks could prove useful in designing antidot-based spin injectors \cite{Zozoulenko2004adspininjector,Dolcetto2013qsh_antidot} or for probing the spin textures of topological insulators \cite{Rod2016qsh_antidot_spintexture}.

\subsection{Network approaches for physical systems}
\label{sec:level2:PhysLaws} 

While network science has its origins in graph theory and sociology -- with many initial applications revolving around the study of social interactions \cite{scott2012social,Wasserman1994Social}, there has been a growing interest in utilizing network-based methods to understand complex physical systems as well. Recent work has used network science techniques for the study of biophysical systems such as structural brain networks \cite{bassett2018nature}, osteocyte networks \cite{Kollmannsberger2017osteocyte,taylor-king2017mean-field}, naturally-occurring flow networks \cite{katifori2012quantifying,shih2015robust,heaton2012analysis}, and multiscale plant biology \cite{duran2017bridging}, to name a few examples. The approach has also proven fruitful in the study of soft-condensed matter systems such as granular materials \cite{papadopoulos2018network}, nanorod dispersions \cite{shi2014network}, and metamaterials \cite{kim2018conformational}. Here, we extend the application of network-based methods further, and examine a system governed not by classical physics, but by quantum physics.

\indent Unlike more traditionally studied networks -- for instance, social networks or the World Wide Web -- the topology and function of networks describing innately physical systems are generally subject to additional constraints that can arise from physical factors. Such factors include the spatial embedding of network elements \cite{barthelemy2011spatial} and the specific physical laws governing the construction and dynamics of and on the network \cite{bianconi2015interdisciplinary,biamonte2017complex}. Thus, accompanying the ongoing use of network science to investigate physical systems, there exists a growing interest in developing physically-motivated network approaches that more explicitly take these constraints and laws into consideration. 

Generalizing network-based methods to incorporate the underlying physics of a networked system is an important step in understanding how various physical rules shape network structure, which should in turn enable an improved understanding of system function and our ability to control or alter that function. Here, we have taken the first steps towards harnessing network-theoretic methods to provide insight into complex many-body quantum phenomena, which is an emerging and exciting area of research. Going forward, network approaches could further illuminate the relationships of energy \cite{Horowitz2015energycost} and entropy \cite{Ben-Shach2013nonabelian_entropy,Hartman2018mesoscopicentropy} to the structure and dynamics of quantum-mechanical states, and provide new metrics to assess their robustness to dissipation and decoherence. This understanding can inform the design and control of quantum devices based on spins \cite{Awschalom2013perspective}, trapped ions \cite{Monroe2013perspective}, superconducting circuits \cite{Devoret2013perspective}, and topological states of matter \cite{Stern2013perspective} for quantum information processing.

\subsection{\label{sec:level2:MethCon} Methodological considerations and future extensions}

There are several theoretical and methodological considerations that are pertinent to this work. First, the transport model that we use is semi-classical in that we assume that the antidot configurations can be approximated using a single-particle picture of occupied spin-orbital states. Second, the sequential transport model does not include lifetime broadening as a result of quantum fluctuations or higher-order co-tunneling processes, both of which may be present in experiments when the energy scale associated with tunneling ($h\Gamma$, where $\Gamma$ is the tunneling rate) is non-negligible compared to the thermal energy. However, in comparing simulations to experiments even in situations where $h\Gamma\approx kT$, we still find very close quantitative agreement \cite{bassett2009probing}. A natural extension of the findings presented here would be to explore the network properties of alternative quantum-mechanical descriptions of the antidot that incorporate interactions and higher-order tunneling processes. Finally, here we consider Rentian scaling as an important topological marker of optimal transport in networks. It could be interesting in the future to consider other markers of transport including navigability \cite{gulyas2015navigable}.\\

\section{\label{sec:level1:Conclusion} Conclusion}

Network science provides a powerful mathematical framework to study the emergent properties of complex quantum systems in a computationally tractable way. Using network science to study the energy state transitions of spin-resolved transport through quantum antidots, we demonstrated that this mesoscopic transport system behaves in a similar manner to other types of biological, computational, and transportation networks. Critically, the nature of the topological Rentian scaling in these systems tracks the complexity of the physical behavior of the system, from unfrustrated to frustrated states. Future work could expand upon these observations to explicitly use the relationship between network topology and physical behavior to effect targeted control of complex semiclassical as well as quantum networks.
	
\section{\label{sec:level1:ack} Acknowledgments}

We thank Christopher W. Lynn and Zhixin Lu for helpful comments on an earlier version of this manuscript. This work was supported by generous funding to D.S.B. and L.C.B from the NSF through the University of Pennsylvania Materials Research Science and Engineering Center (MRSEC) DMR-1720530. L.C.B. would also like to acknowledge support from the Marshall Aid Commemoration Commission and the NSF Graduate Research Fellowship Program. D.S.B. would also like to acknowledge support from the John D. and Catherine T. MacArthur Foundation, the Alfred P. Sloan Foundation, the ISI Foundation, the Paul Allen Foundation, and an NSF CAREER award PHY-1554488. A.N.P. acknowledges support from the Benjamin Franklin Scholars Program and from the University Scholars Program at the University of Pennsylvania. L.P. acknowledges support from the National Science Foundation Graduate Research Fellowships Program. E.T. acknowledges support from the Africk Family. The content is solely the responsibility of the authors and does not necessarily represent the official views of any of the funding agencies.

%\section{\label{sec:level1:Refs} References}
%\bibliographystyle{unsrt}
\bibliography{Poteshman_refs}

\end{document}

\title{Supplementary Materials for\\ ``Network architecture of energy landscapes in mesoscopic quantum systems''}% Force line breaks with \\
\author{Abigail N. Poteshman}
\affiliation{Department of Physics and Astronomy, University of Pennsylvania, Philadelphia, PA 19104, USA}
\author{Evelyn Tang}
\affiliation{Department of Bioengineering, University of Pennsylvania, Philadelphia, PA 19104, USA}
\author{Lia Papadopoulos}
\affiliation{Department of Physics and Astronomy, University of Pennsylvania, Philadelphia, PA 19104, USA}
\author{Danielle S. Bassett}
\email[To whom correspondence should be addressed: ] {dsb@seas.upenn.edu \& lbassett@seas.upenn.edu}
\thanks{These authors contributed equally.}
\affiliation{Department of Physics and Astronomy, University of Pennsylvania, Philadelphia, PA 19104, USA}
\affiliation{Department of Bioengineering, University of Pennsylvania, Philadelphia, PA 19104, USA}
\affiliation{Department of Electrical \& Systems Engineering, University of Pennsylvania, Philadelphia, PA 19104, USA}
\author{Lee C. Bassett}
\email[To whom correspondence should be addressed: ]{dsb@seas.upenn.edu \& lbassett@seas.upenn.edu}
\thanks{These authors contributed equally.}
\affiliation{Department of Electrical \& Systems Engineering, University of Pennsylvania, Philadelphia, PA 19104, USA}

\maketitle

\noindent \textbf{Summary of the Contents of this Supplement.} In this supplementary materials document, we provide additional methodological details and supplementary results that support the findings reported in the main text. We divide the document into three main sections: Computational Model for Quantum Transport (Sec.~\ref{sec:transport_model}), Supplementary Methods (Sec.~\ref{sec:suppl_methods}), and Supplementary Results (Sec.~\ref{sec:suppl_results}). In the Computational Model for Quantum Transport section, we begin in Sec.~\ref{sec:MasterEqn} by describing the calculations that we performed to estimate transition rates for the master equation. In Sec.~\ref{sec:CalcIG}, we describe the calculations that we performed to estimate current and conductance based on the occupation probabilities obtained from the master equation. In Sec.~\ref{sec:TransportModel}, we describe how we modify the model to include spin-dependent selection rules and to calculate spin-resolved measurements of current and conductance. We begin the Supplementary Methods section in Sec.~\ref{sec:thresh} by considering the effect of thresholding the transition probability matrices and further motivating the choice of threshold used to obtain the results reported in the main manuscript. In Sec.~\ref{sec:est_top_Rent}, we describe our method for estimating the value of the topological Rent's exponent. In Sec.~\ref{sec:outliers}, we further describe our method for measuring the statistical significance of topologically complex outliers. Next we turn to a Supplementary Results section beginning in Sec.~\ref{sec:comp_mod} where we report the main findings obtained from a formal model comparison analysis for the partition data. Finally, in Sec.~\ref{sec:phys_sign} we provide additional results derived from a broader assessment of the physical significance of network statistics. Collectively, these additional methodological details and supplementary results serve to support our main findings, and broaden our intuitions regarding their sensitivity, specificity, and robustness. 

\newpage
\section{Computational Model for Quantum Transport}
\label{sec:transport_model}

In this section we provide additional information regarding the computational model for quantum transport used in the work described in the main text. We begin in Sec.~\ref{sec:MasterEqn} by describing the calculations that we performed to estimate transition rates for the master equation. In Sec.~\ref{sec:CalcIG}, we describe the calculations that we performed to estimate current and conductance based on the occupation probabilities obtained from the master equation. In Sec.~\ref{sec:TransportModel}, we describe how we modify the model to include spin-dependent selection rules and to calculate spin-resolved measurements of current and conductance. 

\subsection{Calculating transition rates for the master equation model\label{sec:MasterEqn}}

We describe the quantum mechanics of an antidot using a set of `fermionic' states $\lvert s\rangle =
\lvert \lbrace n_{\ell,\sigma}\rbrace\rangle$, labeled by occupation numbers $n_{\ell,\sigma}=0,1$ for the state with orbital and spin quantum numbers $\ell$ and $\sigma$. We split the system into three parts such that the total Hamiltonian $H = H_\mathrm{antidot} + H_\mathrm{res} + H_\mathrm{tun}$ represents the physics of the antidot, the reservoirs, and the tunneling between them, respectively. The Hamiltonian for the antidot is
\begin{equation}
  H_\mathrm{antidot} = \sum_s E_s \lvert s\rangle\langle s\rvert,
\end{equation}
where, within the constant interaction model,
\begin{equation}
  E_s = \sum_{\ell\sigma}\varepsilon_{\ell\sigma}n_{\ell\sigma} +
  \frac{1}{2}\Ec(N-\nG)^2.
\end{equation}
Here, $\Ec = e^2/C$ is the charging energy determined using the equivalent capacitor network (see Fig.~1C in the main text), $N$ is the number of electrons in the antidot, and $\varepsilon_{\ell\sigma}$ represents the single-particle eigenenergies of the antidot.

Similarly, the Hamiltonians describing the reservoirs and tunneling to and from the antidot are given in second-quantized form by
\begin{subequations}
\begin{align}
  H_\mathrm{res} & = \sum_{r = \mathrm{S,D}}\left[
      \sum_{k\sigma}\varepsilon_\ksr a_\ksr^\dag a_\ksr + \mu_r
      \hat{n}_r \right], \\
  H_\mathrm{tun} & = \sum_{r=\mathrm{S,D}}\left[\sum_\kls
      T^r_\kls a^\dag_\ksr a_\ls + \text{h.c.}\right],
\end{align}
\end{subequations}
where the leads (assumed to be non-interacting) are labeled by reservoir $r$, wave vector $k$, and spin $\sigma$. The operators $a_\ksr$ and $a_\ls$ annihilate particles in the lead states $\lvert k\sigma\rangle$ of reservoir $r$ and antidot states $\lvert\ls\rangle$, respectively, and $\hat{n}_r$ is the particle-number operator for
lead $r$, with chemical potential $\mu_r= -eV_r$.

Assuming the couplings to the leads $T^r_\kls$ are small relative to the thermal energy, $k_\mathrm{B}T$, such that thermal fluctuations dominate over quantum-mechanical fluctuations, we can use Fermi's golden rule to write the tunneling rates for the transition between antidot states $\sptos$ and reservoir states $\kptok$ to first order as
\begin{equation}\label{eq:FGRrate}
  W^p_{s^\prime\chi^\prime\rightarrow s\chi}\simeq
      \frac{2\pi}{\hbar}\Bigl\lvert\langle \chi s\rvert H_\mathrm{tun}
          \lvert \chi^\prime s^\prime\rangle\Bigr\rvert^2
      \delta(E_s - E_{s^\prime} + E_\chi - E_{\chi^\prime}+p\mu_r),
\end{equation}
where $p = \pm1$ denotes the change of electron number on the dot, and $E_\chi$ is the energy of the reservoir state $\chi$. We are interested in the rates between individual antidot states, which are obtained by summing out the contributions from all lead states,
\begin{equation}\label{eq:sumoutleads}
  \gamma^p_\sptos = \mspace{-18mu} \sum_{\substack{\chi\chi^\prime \\ N(\chi^\prime) =
  N(\chi)+p}}
      \mspace{-18mu} W^p_{s^\prime\chi^\prime\rightarrow s\chi}
      \rho^\mathrm{eq}_\mathrm{res}(\chi^\prime),
\end{equation}
where $\rho^\mathrm{eq}_\mathrm{res}$ is the equilibrium density of states in the reservoirs. We obtain the result
\begin{subequations}\label{eq:gammapm}
  \begin{align}
    \gamma^+_{r,\sptos} & = \sum_\llps \Gamma^r_\llps(E_s-E_\sp)
        \bra{s}a^\dag_\ls\ket{\sp} \bra{\sp} a_\lps \ket{s}
        f_r(E_s - E_\sp), \label{eq:gammap}\\
    \gamma^-_{r,\sptos} & = \sum_\llps \Gamma^r_\llps(E_\sp - E_s)
        \bra{s}a_\ls\ket{\sp} \bra{\sp}a^\dag_\lps\ket{s}
        \Bigl[1-f_r(E_\sp - E_s)\Bigr],\label{eq:gammam}
  \end{align}
\end{subequations}
where the spectral function is defined as
\begin{equation}\label{eq:spectralfn}
  \Gamma^r_\llps(E) = \frac{2\pi}{\hbar}\sum_k T^r_\kls
  T^{r\ast}_{k\ell^\prime\sigma} \delta(E-\varepsilon_\ksr),
\end{equation}
and
\begin{equation}
  f_r(E) = \frac{1}{1+e^{(E-\mu_r)/k_\mathrm{B}T}}
\end{equation}
are the Fermi functions that describe the occupation of states in the reservoirs.

\subsection{Calculating current and conductance\label{sec:CalcIG}}

Once the master equation has been solved for the probabilities $P(s)$, we can compute the current flowing out of each lead from the expression
\begin{equation} \label{eq:Ir1}
\begin{split}
  I_r & = e\sum_{s\sp} \bigl[ \gamma^+_{r,\sptos}P(\sp)
      - \gamma^-_{r,s\rightarrow\sp}P(s)\bigr] \\
      & = e\sum_{s\sp}\bigl[\gamma^+_{r,\sptos} -
      \gamma^-_{r,\sptos}\bigr] P(\sp),
\end{split}
\end{equation}
where we have used Eq. (1) in the main text to simplify the second term. Using the relation
\begin{equation}
  \sum_r\bigl[\gamma^+_{r,\sptos} -
      \gamma^-_{r,\sptos}\bigr] = \Bigl(N(s)-N(\sp)\Bigr)\gamma_{\sp
      s},
\end{equation}
it is straightforward to show that $\sum_r I_r = 0$, i.e.\ that the total current is conserved. Dropping the dependence on $\ket{\ls}$, we can write \eqnref{eq:Ir1} in the form
\begin{equation}\label{eq:Ir2}
  I_r = e\sum_{s\sp}\Gamma_r(\mussp) M_{s\sp}\Bigl[f_r(\mussp)P(\sp)
  - \bigl(1-f_r(\mussp)\bigr)P(s)\Bigr],
\end{equation}
where
\begin{equation}
  M_{s\sp} = \sum_\ls
  \bigl\lvert\bra{s}a^\dag_\ls\ket{\sp}\bigr\rvert^2
\end{equation}
represents the selection rules for transitions between states
$s^\prime\leftrightarrow s$ and $\mussp = E_s - E_\sp$ is the chemical potential for the antidot transition. Using current conservation we can derive the relation
\begin{equation}
  \sum_{s\sp}M_{s\sp}P(s) = \sum_{s\sp
  r}\frac{\Gamma_r}{\Gamma}M_{s\sp}\bigl[ P(\sp)+P(s)\bigr]
  f_r(\mussp),
\end{equation}
where $\Gamma = \sum_r \Gamma_r$ and we have suppressed the dependence of the $\Gamma$'s on $\mussp$. We use this relation to eliminate the term independent of $f_r$ in \eqnref{eq:Ir2} to obtain a final expression for the current out of lead $r$,
\begin{equation}
  I_r = e\sum_{s\sp r^\prime}
  \frac{\Gamma_r\Gamma_{r^\prime}}{\Gamma}
      M_{s\sp}\bigl[P(\sp)+P(s)\bigr]
      \times\bigl[f_r(\mussp) - f_{r^\prime}(\mussp)\bigr].
\end{equation}
We use this expression to calculate the current transmitted through the antidot, and compute the conductance at finite bias by
\begin{equation}\label{eq:Conductance}
  G(\Vd) = \frac{I(\Vd + \delta\Vd) - I(\Vd -
  \delta\Vd)}{2\delta\Vd},
\end{equation}
which is typically a good approximation if $e\delta\Vd\lesssim kT$.

\subsection{Model of spin-dependent nonlinear transport \label{sec:TransportModel}}

For a given experimental configuration\,---\,consisting of a set of tunnel barriers (possibly spin- and/or energy-dependent) and bias (possibly mode-dependent through the non-equilibrium population of edge modes)\,---\,we have a `shell' within which we can explore different physical models for the antidot. Given a set of values for the external fields (gate voltages, magnetic field, and drain-source bias), we can determine the ground-state configuration, but if the energy spacing between states is small, or if $\Vds$ is large, the steady-state solution will contain significant populations of many excited states as well. To determine which subset of excited states participates in transport, we start with a relatively small subset of states, chosen to be the ground-state configuration plus all of the excited states that are accessible through a single tunneling event, i.e., the states with energy $\varepsilon_i$ such that the chemical potential
\begin{equation}
  \mu_i = \varepsilon_i - \varepsilon_\mathrm{GS}
\end{equation}
is within the energy window defined by the chemical potentials of the leads:
\begin{equation}
\bigl[\min(\muS,\muD)-\Etherm,\,\max(\muS,\muD)+\Etherm\bigr],
\end{equation}
where $\Etherm\approx 4kT$. If we find that many of these excited states have significant occupation probabilities, we can add to this set all of the states that are connected to the significant excited states through the same rule, in terms of the chemical potentials for transitions from each excited state. By continuing to expand the set of states in this way, we will eventually reach a situation in which all of the newly-added states have sufficiently low occupation probability for convergence of the transport current to a desired tolerance. In the results reported here and in the main text, the occupation probability threshold was $1 \times 10^{-6}$.\\

\indent To calculate the spin-dependent current, we need a method of organizing the antidot configurations that allows us to keep track of the \emph{spin} of each electron which tunnels into or out of the antidot. Here we outline the procedure that we use to accomplish this, using the fermionic configurations defined by occupation vectors $(\nupvec,\ndnvec)$ as an example. To begin, we consider only transitions between ground-state configurations with different occupation numbers $N$ at zero bias, with chemical potentials $\mu_0(N)$. Given a set of capacitances as described in Sec. IIA in the main text, the condition $\mu_0(N)=0$ defines the value of the gate voltage $\Vg$ at which charge degeneracy occurs for the $N\leftrightarrow\Npone$ transition at zero bias. In between these resonance positions, the condition $\mu(\Nmone) = -\mu(N)$ defines an inflection point within each Coulomb blockade region. On one side of the inflection point we need only consider configurations with occupation numbers $(\Nmone,N)$, while on the other we consider only $(N,\Npone)$ states. In the plane of $(\Vg,\Vds)$ these become inflection lines that pass vertically through the center of each Coulomb diamond, and divide the calculation region by the occupation numbers involved.  

Next, we divide the configurations within each region (defined by occupation numbers $N$, $\Npone$) by their total spin projection $S_z$. Suppose the ground-state spin for the $N$-particle state is $\Szz$ and for the $\Npone$ particle state is $\Szz-\frac{1}{2}$. Given these values, we begin by constructing the vector of configurations:
\begin{equation}
    \lbrace{\ket{\Psi_\mathrm{AD}}\rbrace} =
  \begin{pmatrix}
    \lbrace\ket{\Npone,\Szz\!-\!\frac{3}{2}}\rbrace \\
    \lbrace\ket{N,\Szz\!-\!1}\rbrace \\
    \lbrace\ket{\Npone,\Szz\!-\!\frac{1}{2}}\rbrace \\
    \lbrace\ket{N,\Szz}\rbrace \\
    \lbrace\ket{\Npone,\Szz\!+\!\frac{1}{2}}\rbrace \\
    \lbrace\ket{N,\Szz\!-\!1}\rbrace
  \end{pmatrix},
\end{equation}
where each $\lbrace\ket{N,S_z}\rbrace$ corresponds to a vector of individual states $\ket{N,S_z,q_\uparrow,q_\downarrow}$, where $q_\sigma$ labels the configuration of the spin-$\sigma$ particles. In the presence of interactions, these states are not true eigenstates of the Hamiltonian, but they provide a qualitative approximation to the excitation gaps in most cases. 
  
The number of excited states to include is determined through a consideration of the chemical potentials for transitions to or from the ground states with spin $S_z\pm\frac{1}{2}$ as described above. With this arrangement for the configurations, the selection rules take the block-matrix form
\begin{equation}\label{eq:Selrules}
  \begin{pmatrix}
    0 & W^{+\downarrow}_{\Szz-1} & & & \cdots & 0 \\
    W^{-\downarrow}_{\Szz-\frac{3}{2}} & 0 & W^{-\uparrow}_{\Szz-\frac{1}{2}} & & & \vdots \\
     & W^{+\uparrow}_{\Szz-1} & 0 & W^{+\downarrow}_{\Szz}  & & \\
     & & W^{-\downarrow}_{\Szz-\frac{1}{2}} & 0 & W^{-\uparrow}_{\Szz+\frac{1}{2}} & \\
     \vdots & & & W^{+\uparrow}_{\Szz} & 0 & W^{+\downarrow}_{\Szz+1} \\
     0 & \cdots & & & W^{-\downarrow}_{\Szz+\frac{1}{2}} & 0 \\
  \end{pmatrix},
\end{equation}
where, assuming the vectors of states $\lbrace\ket{N,S_z}\rbrace$ are listed as subsequent groups of \spinup\ states (labeled by $q_\uparrow$) for each \spindn\ state (labeled by $q_\downarrow$), the sub-matrices $W^{\pm\sigma}_{S_z}$ are given by
\begin{subequations}
\begin{align}
  W^{\pm\uparrow}_{S_z} & = \mathbf{1}_\downarrow \otimes M^{\pm\uparrow}_{S_z}, \\
  W^{\pm\downarrow}_{S_z} & = M^{\pm\downarrow}_{S_z} \otimes
  \mathbf{1}_\uparrow.
\end{align}
\end{subequations}
The matrices $M^{\pm\sigma}_{S_z}$ contain the selection rules for transitions in the spin-$\sigma$ configuration individually, and are easily worked out by comparing the occupation vectors $\mathbf{n}_\sigma$ of the initial and final states. For example, $M^{+\uparrow}_{ij}=1$ whenever the $q_\uparrow=i$ state of the $\Npone$ configurations results from adding a single \spinup\ particle to the $q_\uparrow=j$ state of the $N$ configurations, which we can write as
\begin{equation}
  M^{+\uparrow}_{ij} = \begin{cases}
    1\quad\text{if}\quad
    \mathbf{n}_\uparrow(i)\cdot[\mathbf{1}-\mathbf{n}_\uparrow(j)]=1,\\
    0\quad\text{otherwise}.
  \end{cases}
\end{equation}
Similar relations determine the selection rules for other types of processes.

\begin{figure}
	\begin{center}
	\includegraphics[width=0.7\textwidth]{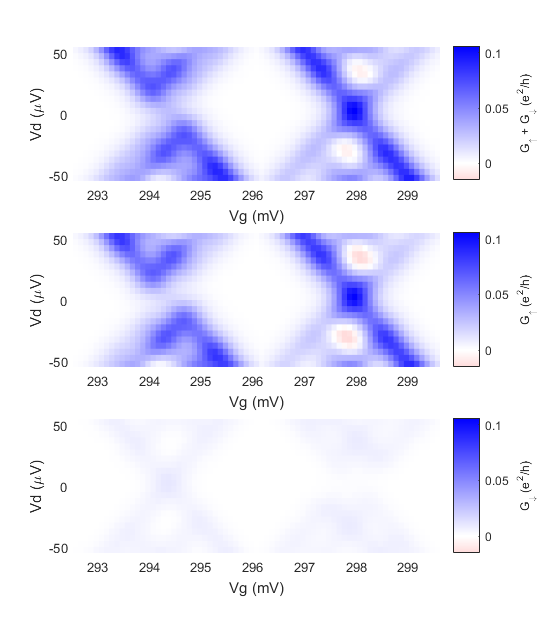}
    \caption{\textbf{Spin-resolved transport simulations.} The total conductance (top) is decomposed into components representing the tunneling of spin-up (middle) and spin-down (bottom) electrons, respectively. The simulation parameters are the same as those used for Fig.~4\&5 of the main text and listed in Sec. IIB. The difference in magnitude between the two spin components reflects spin-dependent tunnel couplings that are chosen to match experimental conditions \cite{bassett2009probing}. Here, $\gSup=1500$~MHz, $\gDup=750$~MHz, $\gSdn=550$~MHz, and $\gDdn=55$~MHz.  }
    \label{fig:SpinResolvedTransport}
    \end{center}
\end{figure}

The rate matrix has a similar form to \eqnref{eq:Selrules}, where the nonzero selection rules are replaced by the transition rates
\begin{equation}
  R^{\pm\sigma}_{ij} =
  \sum_{r=\mathrm{S,D}}\Gamma^r_\sigma(\mu_{ij})W^{\pm\sigma}_{ij}f_r^\pm(\mu_{ij}),
\end{equation}
where $f^+_r = f_r$ is the Fermi function of lead $r$, and $f^-_r = 1-f_r$. As described in Sec. IIC in the main text, we then add diagonal elements to the rate matrix to impose a net balance of rates in equilibrium, and an extra row of ones to enforce normalization, constructing the master equation in the form of Eq. (2) in the main text. The solution to this master equation gives the steady state occupation probability of each state $\ket{N,S_z,q_\uparrow,q_\downarrow}$, which we then use to compute the current flowing through the system. The current is most easily computed using \eqnref{eq:Ir1}, by isolating the transition rate involving only a single lead, e.g.\ for the source,
\begin{equation}
  S^{\pm\sigma}_{ij}=\GammaS_{\sigma}(\mu_{ij})W^{\pm\sigma}_{ij}f_\mathrm{S}^\pm(\mu_{ij}).
\end{equation}
Including signs to account for the direction of current flow, we can then write
\begin{equation}
  I = e\sum_{ij}T_{ij}P_j,
\end{equation}
where $P_j$ are the equilibrium occupation probabilities and
\begin{equation}
  T = \begin{pmatrix}
    0 & S^{+\downarrow}_{\Szz-1} & & & \cdots & 0 \\
    -S^{-\downarrow}_{\Szz-\frac{3}{2}} & 0 & -S^{-\uparrow}_{\Szz-\frac{1}{2}} & & & \vdots \\
     & S^{+\uparrow}_{\Szz-1} & 0 & S^{+\downarrow}_{\Szz}  & & \\
     & & -S^{-\downarrow}_{\Szz-\frac{1}{2}} & 0 & -S^{-\uparrow}_{\Szz+\frac{1}{2}} & \\
     \vdots & & & S^{+\uparrow}_{\Szz} & 0 & S^{+\downarrow}_{\Szz+1} \\
     0 & \cdots & & & -S^{-\downarrow}_{\Szz+\frac{1}{2}} & 0 \\
  \end{pmatrix}.\label{eq:Tmatrix}
\end{equation}
The block-diagonal form of Eq.~(\ref{eq:Tmatrix}) also facilitates the straightforward calculation of spin-resolved current components, simply by isolating the terms that correspond to tunneling of each spin species. The spin-resolved conductance is calculated as in Eq.~(\ref{eq:Conductance}) from a finite difference of the currents computed at two different settings for $\Vd$. Fig.~\ref{fig:SpinResolvedTransport} shows the spin-resolved conductance components corresponding to the simulations used in Fig.~4\&5 of the main text.

The procedure can be generalized to account for additional effects. For example, we can include spin-conserving relaxation of excited states within each set $\lbrace\ket{N,S_z}\rbrace$ by adding block matrices describing these processes to the main diagonal of \eqnref{eq:Selrules}. Spin non-conserving relaxation due to spin-orbit coupling or the hyperfine interaction could also be included by adding terms to the next off-diagonal blocks (connecting states $\lbrace\ket{N,S_z}\rbrace$ with
$\lbrace\ket{N,S_z\!\pm\!1}\rbrace$). Note, however, that this model only obtains the steady-state ($t\rightarrow\infty$) configuration, and it assumes a Markovian (\textit{i.e.}, history-independent) interaction with the reservoirs; thus, we are not able to investigate coherent effects due to quantum evolution with this procedure.

We iterate this procedure until convergence is reached, adding additional $S_z$-configurations and excited states until the occupation probability of the outermost states falls below a predetermined threshold. In the results reported here and in the main text, the total occupation probability threshold of the outermost states was constrained to be less than 0.02. When performing simulations over a range of different parameters as in Fig.~\ref{fig:SpinResolvedTransport}, the matrix-construction procedure is followed independently for each bias setting, and therefore the set of states included in the calculation varies. When investigating the role of some network measures, it is important to maintain a constant network size (\textit{i.e.}, the total number of states) and ideally the same state definitions. Therefore, after performing the calculation once over the full parameter space of interest, we determine the union of all quantum states that appear at any point and subsequently repeat the calculation at each bias point using the full set. A drawback of this approach is that, at every bias point, a large number of states have negligible occupation probability and do not contribute to the dynamics. When appropriate, the non-participatory states can be removed using a thresholding procedure as described in the next section.

\newpage
\section{Supplementary Methods}
\label{sec:suppl_methods}

In this section we provide additional information regarding the details of the methods employed in the work described in the main manuscript, as well as motivations for various methodological choices. We begin in Sec.~\ref{sec:thresh} by considering the effect of thresholding the transition probability matrices and further motivating the choice of threshold used to obtain the results reported in the main manuscript. In Sec.~\ref{sec:est_top_Rent}, we describe our method for estimating the value of the topological Rent's exponent. In Sec.~\ref{sec:outliers}, we further describe our method for measuring the statistical significance of topologically complex outliers.

\subsection{Thresholding probability values}
\label{sec:thresh}

\begin{figure}
	\begin{center}
	\includegraphics[width=0.7\textwidth]{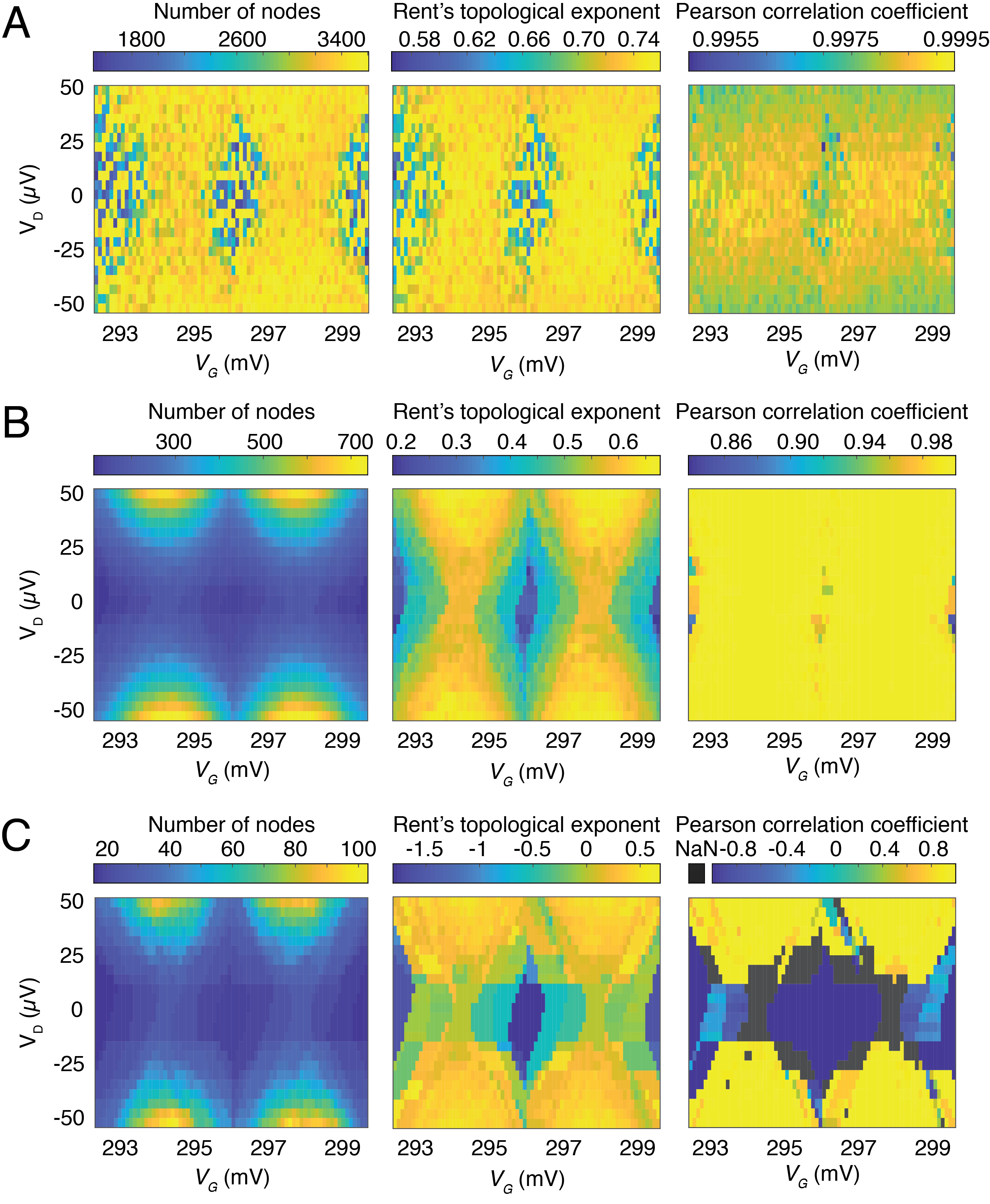}
    \caption{\textbf{Network size, estimates of Rent's topological exponent, and goodness of fit obtained with different values of the threshold on transition probabilities at 55 mK.} \textbf{A.} Values of the number of nodes (left panel), the topological Rent's exponent (middle panel), and the Pearson correlation coefficient (right panel) when a threshold is applied to remove all transition probability values below zero. The values displayed for the Rent's topological exponent and the Pearson correlation coefficient are averages over 10 trials rather than 50. \textbf{B.} Values of the number of nodes (left panel), the topological Rent's exponent (middle panel), and the Pearson correlation coefficient (right panel) when a threshold is applied to remove all transition probability values below $2.9274 \times 10^{-10}$. \textbf{C.} Values of the number of nodes (left panel), the topological Rent's exponent (middle panel), and the Pearson correlation coefficient (right panel) when a threshold is applied to remove all transition probability values below $2.9274 \times 10^{-4}$. NaN is displayed for Pearson correlation coefficient values estimated from networks where the standard deviation of the number of nodes or the number of edges for a set of partition data is zero.}
    \label{fig:ProbThresh}
    \end{center}
\end{figure}

It is important to note that the rate matrix inversion calculations in our model of sequential transport through an antidot produced numerical inaccuracies in probability values near zero. There are several potential ways to deal with these inaccuracies. Intuitively, if we threshold out all probability values less than zero, then all of the probability values that we remove are certainly introduced as a result of numerical inaccuracies (Fig.~\ref{fig:ProbThresh}A). However, without also thresholding out very small positive probability values, we have not necessarily excluded all of the probability values that arise from the numerical inaccuracies. We expect that the majority of the very small positive probability values that are of the same magnitude as the negative probability values are also a result of numerical inaccuracies. To maximize our confidence in the network architecture represented by the transition probability matrix, we therefore wish to find a threshold that simultaneously excludes these very small, positive probabilities and the very small, negative probabilities. 

If we threshold out energy states with a probability below an even probability distribution based on the number of energy states that we are examining, we find that the threshold excludes information that should be preserved in the network model. We define our transition rate matrices to consider 3416 different possible energy states between which the antidot system can transition. Using an even probability distribution, when we exclude probability values below $\frac{1}{3416} \approx 2.93 \times 10^{-4}$, the networks representing accessible energy states were split into multiple connected components; see Fig.~\ref{fig:ProbThresh}C for results when we apply a threshold based on an even probability distribution to our probability values. This solution is inherently unphysical because without an edge for the antidot system to transition from energy states in one connected component to energy states in another connected component, the system remains constrained to a subset of energy states. To ensure that such an unphysical representation is not employed, we seek a threshold that is low enough that the network representing accessible energy states contains a single connected component for the network over voltage settings, resulting in an allowable current through the antidot system; see Fig.~\ref{fig:ProbThresh}B. 

For the majority ($ > 90$\%) of the networks, the threshold employed preserves the connectedness of networks corresponding to voltage settings under which the antidot system experiences a current greater than 1 pA. For fewer than 10\% of the networks, this threshold resulted in multiple connected components, but the distribution of nodes and occupation probabilities across connected components was not even. Specifically, this threshold resulted in one connected component where the probability that the antidot system would be in an energy state represented by a node in this single connected component was greater than $0.999999995$. The smaller connected components consisted of fewer than 8\% of the number of nodes after the threshold was applied were removed so that only this single connected component remained. All of the network analyses presented in the paper were performed on networks where the smaller connected components were removed if the network was split into multiple connected components after the threshold was applied.

\subsection{Estimating the value of the topological Rent's exponent}
\label{sec:est_top_Rent}

\begin{figure}
	\begin{center}
	\includegraphics[width=0.50\textwidth]{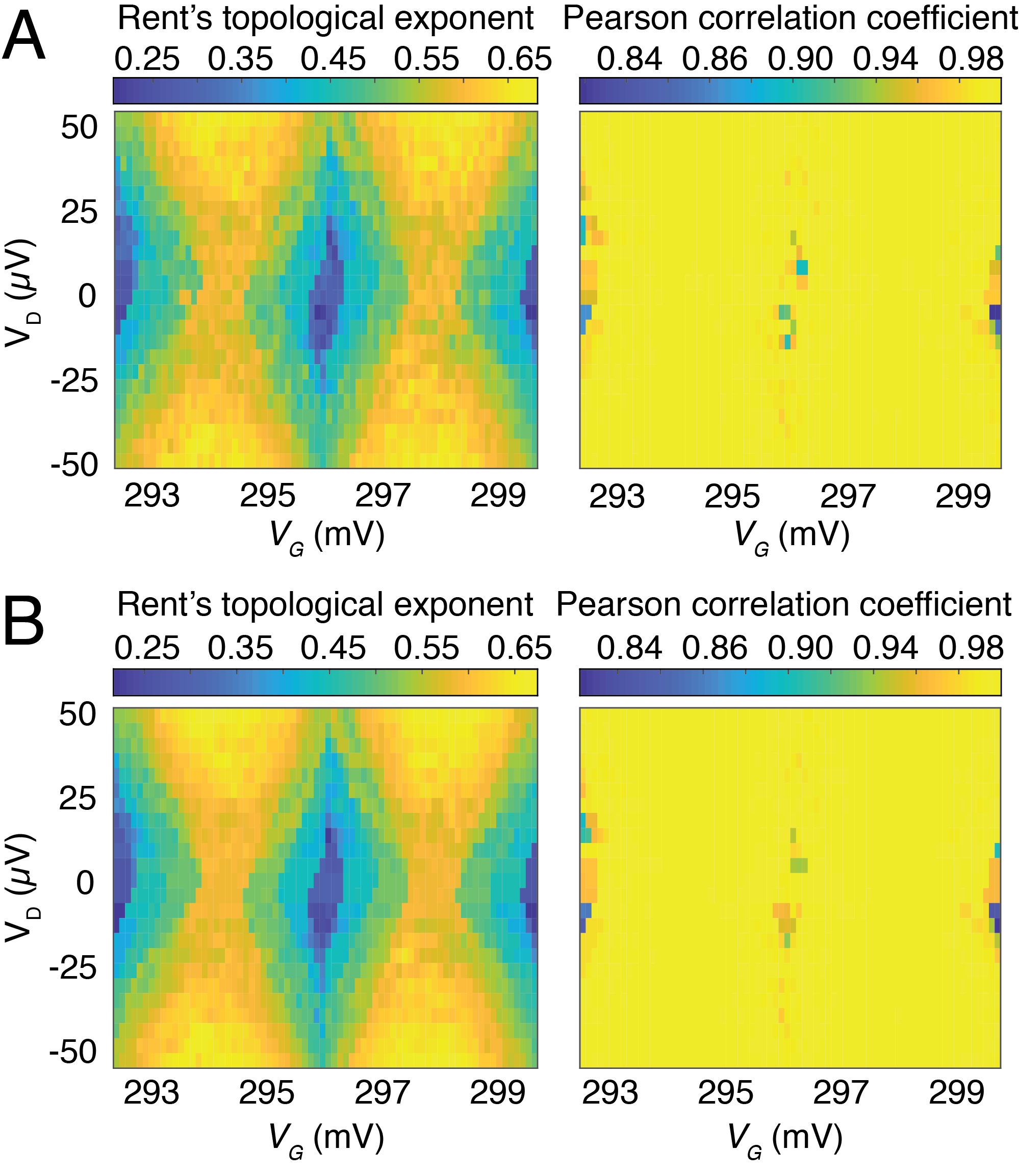}
    \caption{\textbf{Correspondence between single-run estimates and the estimate obtained by averaging across 50 runs of the heuristic bipartitioning algorithm.} \textbf{A.} Values of the topological Rent's exponent (left panel) and the Pearson correlation coefficient (right panel) from one trial of partition data. \textbf{B.} Values of the topological Rent's exponent (left panel) and the Pearson correlation coefficient (right panel) obtained from averaging estimates over 50 trials of partition data.}
    \label{fig:AvgEst}
    \end{center}
\end{figure}

\indent To test for the presence of topological Rentian scaling, we used a recursive bipartitioning algorithm (hyper-graph partitioning package \textit{hMETIS}, version 1.5.3 \cite{karypis1999multilevel}) that minimizes the number of edges crossing a partition boundary for each cut. The \textit{hMETIS} bipartitioning algorithm is a non-deterministic heuristic, and therefore different runs of the algorithm yield slightly different values of the topological Rent's exponent. To account for this variation in the estimation of the exponent, we ran the partitioning algorithm 50 times for each network. Each run of the partitioning algorithm results in a set of partitioning data, which includes pairs of the number of nodes in a partition and the number of edges crossing the boundary of a partition for different partitions, the sizes of which were determined recursively. For each set of partitioning data, we used MATLAB's linear least squares regression to estimate the topological Rent's exponent as the slope of the best fit line of (i) the log of the number of edges \emph{versus} (ii) the log of the number of nodes. We used the Pearson correlation coefficient to determine the goodness of fit between the same two variables: the log of the number of edges and the log of the number of nodes. We averaged the values of the Pearson correlation coefficient over the 50 trials for each network, and we also averaged the values of the topological Rent's exponent over the 50 trials for each network. See Fig.~\ref{fig:AvgEst} for the correspondence between estimates from a single run of the heuristic bipartitioning algorithm and estimates obtained by averaging across 50 runs. Here and in the main text, we record the averaged values of the topological Rent's exponent and of the Pearson correlation coefficient over 50 trials unless noted otherwise.\\

\subsection{Assessing the statistical significance of differences between frustrated and non-frustrated networks}
\label{sec:outliers}

In the main text, we sought to quantify differences between networks corresponding to frustrated antidots and networks corresponding to non-frustrated antidots that have access to a similar number of energy states. In particular, we considered two different test statistics: \emph{(1)} the difference between the average values of the Rent's topological exponent of the frustrated and control groups ($d_{t}$), and \emph{(2)} the distance between the center of mass of the frustrated and non-frustrated networks, computed in the two-dimensional space defined by the Rent's exponent and the conductance ($d_{t,G}$). For the difference measure, we defined $d_{t}$ such that positive values mean the average Rent's exponent is greater for the frustrated group compared to the non-frustrated control group.

\begin{figure}
	\begin{center}
	\includegraphics[width=\textwidth]{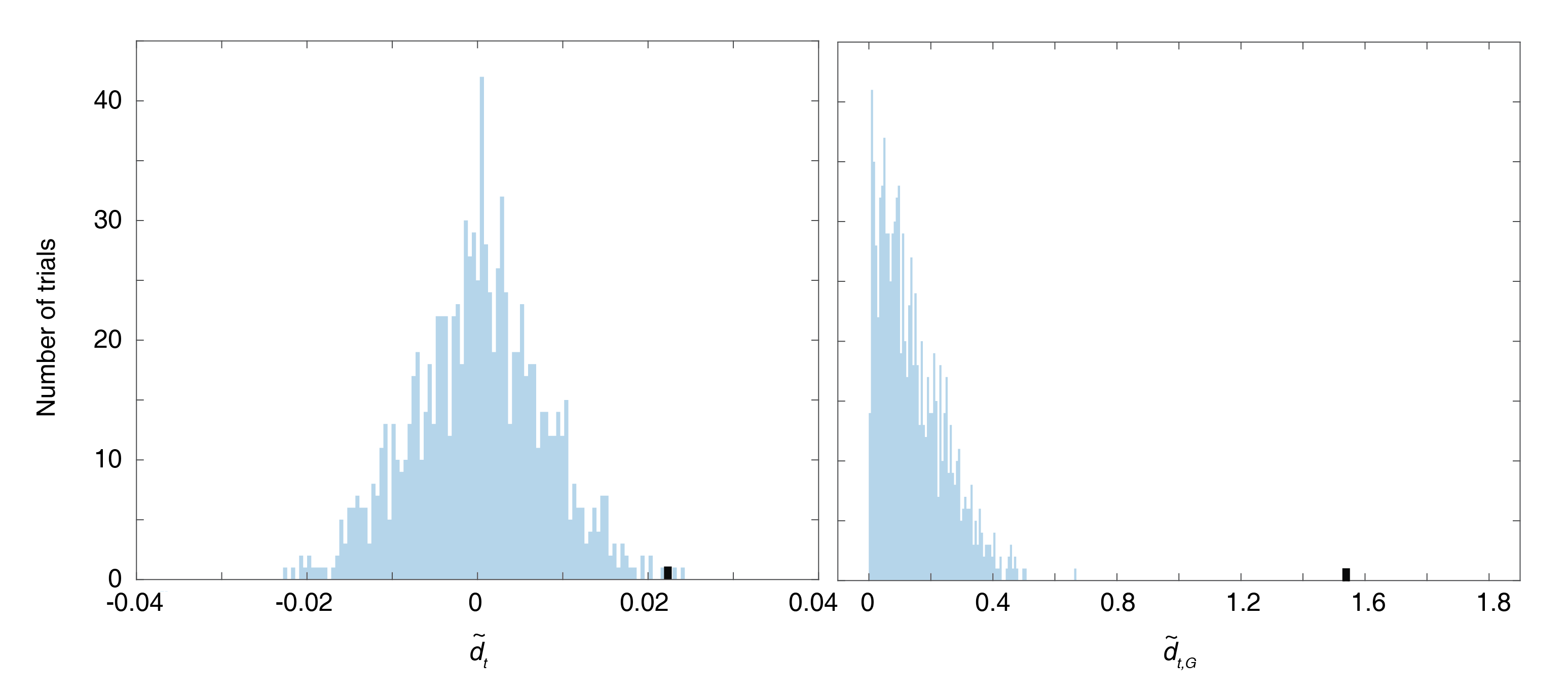}
    \caption{\textbf{Results from permutation tests used to assess the statistical significance of differences between the frustrated and non-frustrated networks at 50 mK.} The left panel shows the null distribution of $\tilde{d}_t$ obtained from 1000 random permutations of the frustrated network and control network labels (blue), and the true value $d_t = 0.0224$ (black). The right panel shows the null distribution of $\tilde{d}_{t,G}$ obtained from 1000 random permutations of the frustrated network and control network labels (blue), and the true value $d_{t,G} = 1.5375$ (black).}
     \label{fig:DistTest}
    \end{center}
\end{figure}

In order to determine whether there are statistically significant separations between the frustrated and control configurations in terms of their positioning along the Rent's exponent axis (which we quantify with $d_{t}$) and their positioning in the plane defined by Rent's exponent and conductance (which we quantify with $d_{t,G}$), we performed nonparametric permutation tests in which the labels --``frustrated" or ``control" -- of the networks are randomized. In particular, for both $d_{t}$ and $d_{t,G}$, we defined two vectors: one with the length of the number of networks corresponding to voltage settings where the antidot experiences frustration, and the other with the length of the number of networks corresponding to voltage settings where the antidot has access to a similar number of energy states but does not experience frustration (i.e, the length of the control set). We then randomly assigned networks to the two vectors and calculated the difference between the average Rent's exponents of the two randomized sets ($\tilde{d}_{t}$) and the distance between the centers of mass of the two randomized groups computed in the two-dimensional space of Rent's exponent and conductance ($\tilde{d}_{t,G}$). The randomization procedure and subsequent calculation of $\tilde{d}_{t}$ and $\tilde{d}_{t,G}$ was carried out 1000 times (see Fig.~\ref{fig:DistTest} for distributions of $\tilde{d}_{t}$ and $\tilde{d}_{t,G}$ at 50mK). Finally, we computed $p$-values for $d_{t}$ and $d_{t,G}$ as the fraction of times $d_{t} < \tilde{d}_{t}$ and the fraction of times $d_{t,G} < \tilde{d}_{t,G}$, respectively.

We found that the non-parametric permutation-based $p$-values for both the $d_{t,G}$ and $d_{t}$ test statistics were statistically significant for all temperatures (all $p$-values $< 0.005$). In terms of $d_{t,G}$, this result suggests a persistent separation -- in the two-dimensional space defined by Rent's topological exponent and conductance -- between the frustrated networks and non-frustrated networks with access to a comparable number of energy states. Furthermore, since $d_t > 0$ and remains statistically significant across all temperatures, we further conclude that the frustrated networks exhibit enhanced topological complexity relative to the control set consisting of non-frustrated networks with access to a comparable number of energy states, and that this difference is robust to changes in the temperature.

%\newpage
\section{Supplementary Results}
\label{sec:suppl_results}

In this section we report the results of additional computations and analyses that complement those that were reported in the main manuscript. We begin in Sec.~\ref{sec:comp_mod} by reporting the main findings obtained from a formal model comparison analysis for the partition data. Then in Sec.~\ref{sec:phys_sign} we provide additional results derived from a broader assessment of the physical significance of network statistics. 

\subsection{Comparing different model forms for the partition data}
\label{sec:comp_mod}

\indent The number of nodes in a partition versus the number of edges crossing the boundary of a partition visually appears linear in log-log space (see Fig. 3D in the main text), thus suggesting the relevance of a power law model for explaining our partitioning data. However, it is important to perform formal model comparison to provide a quantitative (rather than qualitative) basis for selecting a power law model over other comparable models. We therefore tested the power law against quadratic, cubic, exponential, linear, and logarithmic models in order to verify that a power law form is the model most likely to have generated the partition data. Because the number of free parameters varies across different models, we cannot compare the goodness of fit of the models by directly comparing $R^{2}$ values. The Bayesian information criterion, however, provides a criterion for model selection based on the likelihood function and number of data points while simultaneously penalizing the number of free parameters \cite{claeskens2008model}. The model with the lowest Bayesian information criterion is the preferred model. 

We compared models by averaging the $R^2$ values and Bayesian information criterion over 50 trials of partitioning data for each network. We then averaged the $R^2$ and Bayesian information criterion over data from networks corresponding to antidots under voltage settings that experience current; see Table ~\ref{tab:CompModels}. While we found that the cubic model has an average $R^{2}$ value closer to one than the average $R^{2}$ value for the power law model, the Bayesian information criterion is lowest for the power law model. The high value of the Bayesian information criterion for the cubic model indicates that the cubic model is likely over-fitting the partition data. From these results, we conclude that the power law is the most parsimonious model likely to have generated the partition data.

\begin{table}
\begin{center}
\begin{tabular}{ c c c c } 
 \hline
 \textbf{Model} & \textbf{Form} & 
$\mathbf{R^{2}}$ & \textbf{Bayesian information criterion} \\ 
 \hline
 cubic & $y = ax^3 + bx^2 + cx + d$ & $ 0.9942 \pm  \num{1.606e-5}$  & $26.51 \pm 0.018$ \\ 
 exponential & $y = ae^{bx}$ & $0.9779 \pm \num{8.785e-6}$ & $-1.176 \pm 0.002 $\\ 
 linear & $y = ax + b$ &$0.8970 \pm \num{3.015e-5}$& $41.59 \pm 0.002$ \\
 logarithmic & $ y = a\log(x) + b$ & $0.7699 \pm \num{3.410e-5} $& $46.43 \pm 0.001$\\
 power & $y = ax^b$ & $0.9884 \pm \num{7.954e-6}$& $-15.14 \pm 0.004$\\
 quadratic & $y = ax^2 + bx + c$ & $0.9775 \pm \num{2.572e-5} $& $34.04 \pm 0.007$ \\
 \hline
\end{tabular}
\caption{Bayesian information criterion and goodness of fit of different models. Average $R^{2}$ values and Bayesian information criterion values over partition data for networks corresponding to voltage settings where the antidot experiences current.}
\label{tab:CompModels}
\end{center}
\end{table}

\subsection{Assessing the physical significance of network statistics}
\label{sec:phys_sign}

\begin{figure}
	\begin{center}
	\includegraphics[width=0.7\textwidth]{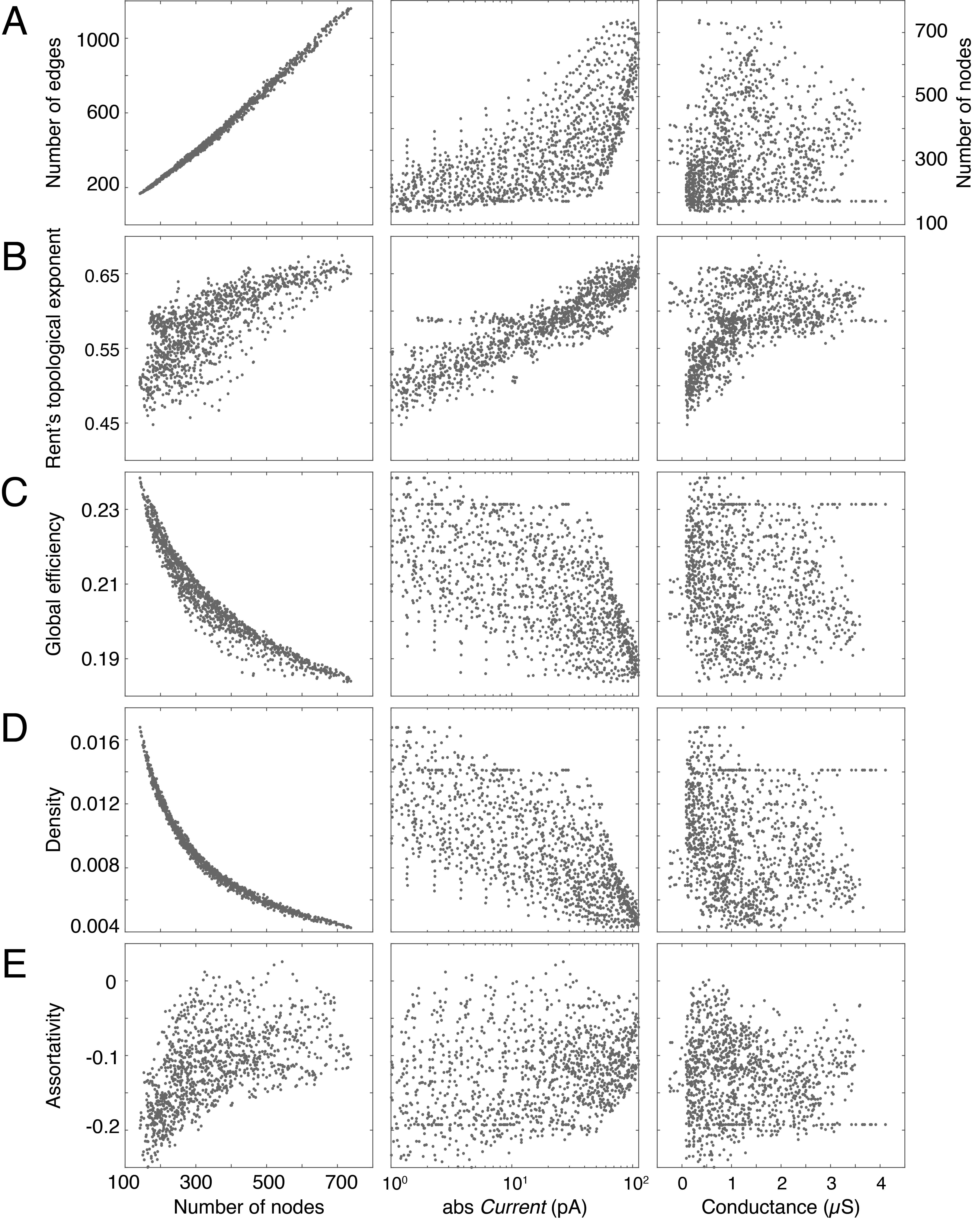}
    \caption{\textbf{Assessing the sensitivity of network statistics to physical properties of the antidot system.} Specifically, we compared the \textbf{A.} number of edges (left panel) or number of nodes (middle and right panels) in the network, \textbf{B.} topological Rent's exponent, \textbf{C.} global efficiency, \textbf{D.} network density, and \textbf{E.} assortativity with the number of nodes (left panel) in the network, the physical parameter of log of the absolute value of current (middle panel) of the antidot system, and the physical parameter of conductance (right panel) of the antidot system. These statistics were measured over network representations of sequential transport through an antidot at 55 mK.}
    \label{fig:CompStats}
    \end{center}
\end{figure}

In using the conceptual frameworks and computational tools from network science to characterize and describe the energy landscape of the antidot system, we seek to ensure that the measure of topological complexity that we employ is sensitive to the underlying physics of the antidot system and insensitive to trivial features dictated by the network's construction. To determine physical sensitivity of our network measures, we sought meaningful relationships between particular network measures and the physical parameters of current and conductance. To determine insensitivity to trivial features, we tested for uninteresting relationships between particular network measures and the size of the network. Here, we define network size by the number of nodes $s$ in the network. The number of nodes $s$ scales directly with the number of edges in the network; see right panel of Fig.~\ref{fig:CompStats}A. Notably, we were able to make this distinction between network size and the physical parameters of current and conductance because the two are not directly related; see middle and left panels of Fig.~\ref{fig:CompStats}A.\\

\indent As stated in Sec.~III in the main text, we observe a scaling relationship between Rent's topological exponent and the log of current as well as with the conductance, whereas we do not observe a similarly clear relationship between Rent's topological exponent and the size of the network; see middle and left panels of Fig.~\ref{fig:CompStats}B. Critically, this physical sensitivity is not shared with other network statistics.\\

\indent Global efficiency is a network measure that quantifies the ease of communication between nodes in a network. Similar to Rentian scaling, global efficiency has been used to study physical transportation and communication networks that have a spatial embedding, including brains \cite{bullmore2009complex} and urban street networks \cite{buhl2006topological,cardillo2006structural}. For a network with $N$ nodes, the average efficiency $E_{avg}$ of a network is the average of the inverse of the shortest path length between all nodes, and global efficiency is the average efficiency normalized by the average efficiency of a fully-connected network with $s$ nodes \cite{latora2001efficient}. Average efficiency is given by
\begin{equation}
E_{avg} = \frac{1}{N(N-1)} \sum_{i \neq j} \frac{1}{d_{ij}}
\end{equation}
where $d_{ij}$ is the shortest topological distance between nodes $i$ and $j$ in the network. Network density is another commonly reported network statistic that indicates how close the network of interest is to a complete or fully-connected network \cite{friedkin1981development}. Network density $D$ is given by 
\begin{equation}
D = \frac{2\epsilon}{N(N-1)}
\end{equation}
where $\epsilon$ is the number of edges in the network and $N$ is the number of nodes. We observe that both network density and global efficiency are closely related to network size; see Fig.~\ref{fig:CompStats}C-D. Furthermore, in contrast to Rent's topological exponent, we do not observe meaningful relationships between these two network measures and the physical parameters of the antidot system; see Fig.~\ref{fig:CompStats}C-D. \\

\indent Two other network measures that we explored are assortativity and average clustering coefficient. Assortativity is a network measure that quantifies correlations between nodes of similar degree \cite{newman2002assortative}. The assortativity coefficient $r$ is the Pearson correlation coefficient of degree between pairs of nodes connected by an edge. Many social networks are assortative, whereas many technological and biological networks are disassortative \cite{newman2002assortative,foster2010edge}. While we observe that many of our networks are disassortative, which further emphasizes the similarity of these energy state transition networks to technological and biological networks beyond the presence of Rentian scaling, we do not observe any interesting relationship between network size or physical parameters and assortativity; see Fig.~\ref{fig:CompStats}E. Because assortativity lacks the same kind of interesting relationship with physical parameters, we suggest that assortativity does not capture aspects of the network topology salient to the underlying physics of the antidot system. The average clustering coefficient can be used to measure how well-connected neighbors of a node are in a graph \cite{soffer2005network}. Interestingly, we observed that the energy state transition networks lack loops consisting of 3 nodes connected by 3 edges, and as a result, we found that the average clustering coefficient for all networks was zero. While outside of the scope of this work, the disassortative nature of the energy state transition networks and the lack of 3-loops suggest that the degree distribution and loop structure in these energy state transitions networks may be interesting to explore further in future work.\\

%\clearpage
%\newpage
%\bibliographystyle{unsrt}
\bibliography{ref_supplement.bib}